\documentclass{JHEP3}
\usepackage{epsfig}

\newcommand{\be}{\begin{equation}}
\newcommand{\ee}{\end{equation}}
\newcommand{\bea}{\begin{eqnarray}}
\newcommand{\eea}{\end{eqnarray}}

\def\IZ{\relax\ifmmode\hbox{Z\kern-.4em Z}\else{Z\kern-.4em Z}\fi}
\newcommand{\non}{\nonumber \\}
\def\half{{1 \over 2}} 
\def\pa{{\partial}}

\def\cU{{\cal U}} \def\cV{{ \cal V}}

\def\hA{\hat{A}}  \def\hC{\hat{C}} \def\hT{\hat{T}} \def\hS{\hat{S}} \def\hL{\hat{L}}
\def\hm{\hat{m}} \def\hn{\hat{n}} \def\htau{\hat{\tau}} \def\hkappa{\hat{\kappa}}

 \def\lam{\lambda}

\newcommand{\sbsection}[1]{\vspace{.5cm} \noindent {\it #1}}

\def\Schw{Schwarzschild }

\def\({\left(} \def\){\right)}
\def\[{\left[} \def\]{\right]}

\preprint{{\tt gr-qc/0608115}}

\title{ \center{Non-uniform black strings in various dimensions}}

\author{Evgeny Sorkin\\
 Department of Physics and Astronomy, University of British Columbia\\
 6224 Agricultural Road, Vancouver,  V6G 1Z1, Canada \\
 {\tt  evgeny@physics.ubc.ca }
}
\date{today}
\abstract{ The nonuniform black strings branch, which emerges from
the critical Gregory-Laflamme string, is numerically constructed  in
dimensions $6\leq D\leq11$ and extended into the strongly non-linear
regime. All the solutions are more massive and less entropic than
the marginal string. We find the asymptotic values of the mass, the
entropy and other physical variables in the limit of large horizon
deformations. By explicit metric comparison we verify that the local
geometry around the ``waist'' of our most nonuniform solutions is
cone-like with less than $10 \% $ deviation. We find evidence that
in this regime the characteristic length scale has a power-law
dependence on a parameter along the branch of the solutions, and
estimate the critical exponent. }

\begin{document}

\section{Introduction and summary}
\label{sec_intro}

The surprising discovery of Gregory and Laflamme (GL) \cite{GL1}
that below certain mass the uniform black strings, existing in the
backgrounds with compact extra dimensions, are perturbatively
unstable created considerable excitement. It initiated intensive
research aimed to understand the consequences of this instability
and to figure out its endpoint.  In this regard, some progress has
been made in the numerical simulations of the decaying string
\cite{CLOPPV}. Yet, the endpoint remains elusive as the simulations
crash (due to numerical problems) when the spacetime is still highly
dynamical and far from settling down to the static solution. A
different program \cite{TopChange} to determine the endstate
proposes to find a phase diagram accommodating all possible static
solutions. The hope is that in principle the endstate can be
inferred  from the diagram since by definition it is a static
solution. This research has recently culminated in the reviews
\cite{Kol_review,HO_review}.

\begin{figure}
\centering
\noindent
\includegraphics[width=12cm]{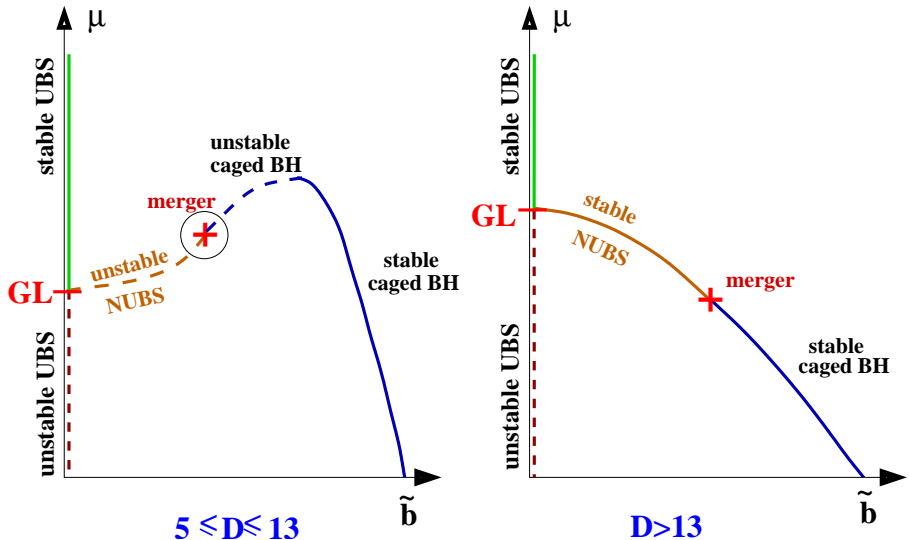}
\caption[]{ A suggested phase diagram. The vertical axis is the dimensionless mass density and the horizontal axis is related to the scalar charge. Shown is the GL point where the uniform string becomes marginally unstable and from which a new branch of non-uniform strings emerges. This branch extends until it meets the caged black hole branch at the ``merger point''. In $D>13$  the NUBSs branch is less massive than the critical string indicating that the order of the phase transition, triggered by appearance of the GL-mode,  changes from first to second.
} \label{fig_phasediagram}
\end{figure}
A proposed phase diagram is portrayed in figure
\ref{fig_phasediagram}. The vertical axis shows the dimensionless
mass density while the horizontal axis is related to the scalar
charge. The branches of solutions represented by solid lines are
presumably stable and the dashed lines indicate unstable solutions.
The diagram has several remarkable features: 1) The merger point
where the black hole branch meets the black string branch and where
the horizon topology changes. 2) A critical dimension $D_{2d~ order}
= 13$ above which the order of the phase transition changes from the
first to the second \cite{CritDim} and where the emergent branch of
nonuniform solutions becomes (thermodynamically) stable
\cite{LGinGL}. According to the diagram, the endpoint of the GL
instability depends on the dimension and it is either a caged black
hole in  $D \leq 13 $ or a nonuniform black string otherwise. For
now, only certain parts of the diagram in certain dimensions were
verified by analytical/numerical constructions. This is described
next.

The black hole branch has been constructed numerically in five
\cite{cagedBHsII,KudohWiseman2} and six
\cite{KudohWiseman2,KudohWiseman1} dimensions. In particular, it has
been shown to approach the merger point, where the north and the
south poles of the caged black hole are expected to intersect across
the compact dimension \cite{KudohWiseman2}.  In other dimensions
only the beginning of the BHs branch was constructed perturbatively
from the Schwarzschild-Tangherlini \cite{Tangherlini} solution in
\cite{Harmark,KolGorbonos1,KolGorbonos2,QFT_BHs} and numerically in
\cite{cagedBHs_unpublished}. The branch of the NUBS was found
perturbatively from the critical GL string in five \cite{Gubser},
six \cite{Wiseman1} and in other, up to sixteen, dimensions
\cite{CritDim}. In six \cite{Wiseman1} and in five
\cite{Oldenburg_BS} dimensions the branch was numerically extended
into non-linear regime, where it was shown to behave in accordance
with diagram \ref{fig_phasediagram}.

In this paper we aim to construct the NUBS branch in higher
dimensions and to extend it beyond linear perturbations into the
strongly non-uniform regime.  To define the ``non-linear'',
``strongly non-uniform'' regime we use a simple geometrical measure
of the horizon deformation \cite{Gubser},
\be
\label{lambda}
\lambda \equiv \half \( {R_{max} \over R_{min}}-1\),
\ee
where $R_{min}/R_{max}$ are the minimal/maximal areal radii of
$z=const$ sections of the horizon. $\lam=0$ for uniform strings and
$\lambda\rightarrow \infty$  at the pinch-off.\footnote{Apparently,
$R_{max}$ remains finite in this limit.} It turns out that  while
at pinch-off, $R_{min} \rightarrow 0$, many other thermodynamical
and geometrical quantities, describing the NUBSs, asymptote to
finite values. This fact suggests a natural definition of the
strongly non-linear regime as a situation when these variables are
close to saturation.\footnote{The saturation is best noted using
$\lambda$, since in $\lam$ the measurables approach their limiting
values roughly exponentially.}

Recently, intriguing proposals regarding the local geometry near the
``waist" of  the extremely non-uniform strings have been put forward
by Kol \cite{TopChange,Kol_on_Choptuik}. Specifically, in
\cite{TopChange} Kol advocates that locally the  merger spacetime is
cone-like, and moreover in $D\leq10$ the local slightly off-merger
metric and its functions have  power-law dependence on parametric
distance from the critical (merger) cone. In other words, if $p$ is
a parametrization of the static NUBSs branch, such that $p*$
corresponds to the merger, than quantities with the length
dimensions are conjectured to scale as
\be
\label{scaling}
\log(\ell) = \gamma \log(p-p*) + \phi\[\gamma \,log(p-p*)+const\] +const,
\ee
where $\phi$ is a periodic function.\footnote{This sort of behavior
is known to appear \cite{HodPiran,Gundlach}  in the near-critical
collapse of a scalar field \cite{Choptuik} (see \cite{OrenSorkin}
for a $D$-dimensional version). Indeed, there are speculations
\cite{Kol_on_Choptuik} that the merger cone might be related to the
critical collapse by a double analytic continuation and change of
the boundary conditions.} Namely the leading behavior is a
power-law with an exponent $\gamma$;  and this is dressed with
periodic ``wiggles".

Unfortunately, the data available in literature provides only
limited insights on what happens at the merger since neither along
NUBS nor along the BH branches is the merger point approached close
enough. Still, some evidence in favor of the local cone geometry was
reported in \cite{KolWiseman}. The scaling (\ref{scaling}) has not
been tested so far.

In this paper, we partially close the gaps. We use numerical
techniques to find fully non-linear branches of NUBSs in dimensions
from six to eleven and confirm that these are in accordance with
diagram \ref{fig_phasediagram}. In dimensions  6, 7, 8, and 9  we
are able to extend the NUBSs branch into the deeply non-linear
regime. In this range of the dimensions we find the limiting values
of the thermodynamic and geometric variables, these are listed in
table \ref{table_variables}. Furthermore, since our method allows
approaching the merger very closely, we compare the local near
``waist" geometry with a cone. This is explicitly done in 6D where
we have our most nonuniform solutions with the highest resolution.
We find that the metrics agree well and the agreement improves with
growing $\lam$. For our most nonuniform solution the discrepancy
between the near-waist metric and the cone does not exceed $10\%$.

Next, we verify the scaling (\ref{scaling}) and find evidence for
the power-law.  We extract the approximate critical exponents and
list them in table \ref{table_gammas}. However, the wiggles are not
seen for our non-uniform solutions. We believe that in order to
actually test for the presence of wiggles, one must have solutions
spanning over several orders of magnitude in $\lam$. Currently, this
challenging task stands beyond the capabilities of our numerical
experiments.

In the next section we describe the setup and derive the field
equations and the boundary conditions. We define the physical
variables and describe how the cone and the scaling (\ref{scaling})
are tested. In section \ref{sec_numerics} we elaborate on our
numerical method. In section \ref{sec_results} we describe the
solutions and their properties and discuss them in section
\ref{sec_discussion}. Various technical details are found in the
appendices.

\section{Static black strings}
\label{sec_setup}

We consider D-dimensional background of the form
$\mathbb{R}^{D-2,1}\times\mathbb{S}^1$, that has one spatial
dimension,  $z$,  compactified on a circle of length L, $z \sim
z+L$. We are  interested in static black objects, which are
spherically symmetric in the extended dimensions. The simplest  such
a solution is the uniform black string whose metric in \Schw
coordinates is given by
\bea
\label{ds_ubs}
ds_{c}^2 &=&ds_{Schw_d}^2 +dz^2,  \non
ds_{Schw_d}^2&=&-f(\rho)dt^2+f(\rho)^{-1} d\rho^2+\rho^2 d\Omega_{d-2}^2,
\eea
where $f(\rho)=1-(\rho_0/\rho)^{d-3}$, $\rho_0$ designates the
horizon location and $d\Omega_{d-2}^2$ is the metric on a unit
sphere $\mathbb{S}^{d-2}$. Gregory and Laflamme  showed that this
string is unstable below certain mass \cite{GL1}. Here, we construct
the branch of static non-uniform black strings that emanates from
the GL-point.

The most general metric describing the black strings can be written as
\be
\label{metric_gen}
ds^2= -e^{2\,\hA} dt^2 + d\sigma(r,z)^2 + e^{2\, \hC} d\Omega_{D-3}^2 ,
\ee
where $\hA$ and $\hC$ are functions of $r$ and $z$, $d\sigma(r,z)^2$
is a two-dimensional metric in the $(r,z)$ plane. The horizon is
located where $\exp(\hA)=0$. A convenient ansatz that we employ in
this paper\footnote{We found that the explicit separation of the UBS
background is essential for stability of our numerical scheme.} uses
``conformal" coordinates in which
\be
\label{metric}
ds^2= - f(\rho(r)) e^{2A} +e^{2 \,B} (dr^2+dz^2) +\rho(r)^2 e^{2C} d\Omega_{D-3}^2 .
\ee
When $A=B=C=0$ we reproduce the UBS solution in conformal
coordinates (the transformation $\rho  \mapsto r $ from \Schw to the
conformal coordinates is described in appendix
\ref{appendix_schw_conf}.)  The horizon is located at the radius
$r_0$, where $f(r_0)=0$. Without loss of generality we assume
$r_0=0$.

\subsection{Equations and boundary conditions}

In order to find the black strings we solve the Einstein equations, $R_{ab} =0$, that
in our coordinates (\ref{metric}) split into three elliptic equations
\bea
\label{Eqs}
\bigtriangleup A &+& \(\pa_r A\)^2  + \(\pa_z A\)^2 +(D-3)\, \( \pa_r A \pa_r C +   \pa_z A \pa_z C \) + \non
&+& {D-4 \over 2 \, \rho\, \sqrt{f}}\(2 \,\pa_r A+ (D-3)\pa_r C \)  + {\sqrt{f} \over 2 \rho}\,\
\( 2 \,\pa_r A -(D-3)(D-4)\, \pa_r C \) =0, \non
\bigtriangleup B &-& {D-3\over 2} \(2\,\pa_r A +(D-4) \pa_r C\) \pa_rC
-{D-3\over 2} \(2\,\pa_z A +(D-4) \pa_z C\) \pa_zC - \non
&-&
 {(D-3)(D-4)\over 2\, \rho \sqrt{f} } \pa_rC -{(D-3) \sqrt{f}\over 2 \,\rho }
 \(2\,\pa_rA+(D-4)\pa_r C \) - \non
 &-&  {(D-3)(D-4) \over 2} {1- e^{2\,B-2\,C} \over \rho^2} =0,\\
\bigtriangleup C &+& \(\pa_r A +(D-3) \pa_r C\) \pa_rC + \(\pa_z A +(D-3) \pa_z C\) \pa_zC + \non
&+&
 {(D-4)\over 2 \,\rho \sqrt{f} } \pa_rC +{ \sqrt{f}\over 2\, \rho } \(2\pa_rA+(3 D-8)\pa_r C \)
 +  (D-4) \, {1- e^{2\,B-2\,C} \over \rho^2} =0 \nonumber,
\eea
where $\bigtriangleup \equiv \pa_z^2  +\pa_r^2 $, and two hyperbolic constraints
\bea
\label{constraints}
G^r_z :  &&\pa_{rz} A + (D-3) \pa_{rz}C -\pa_r A \left(\pa_zB-\pa_zA\right) -  \non
-&& (D-3)\pa_r C \(\pa_zB-\pa_zC\) -\pa_r B \(\pa_z A +(D-3) \pa_z C   \) + \non
+&& {(D-4)\,(1-f) \over 2 \sqrt{f} \,\rho} \( \pa_z A- \pa_z B\) - {(D-3)\,\sqrt{f} \,\over \rho}
 \( \pa_z B -\pa_z C \)=0, \non
G_z^z -G_r^r :&& \pa_z^2 A -\pa_r^2 A +(D-3)\( \pa_z^2 C -\pa_r^2 C \) +
\(\pa_z A \)^2 -\(\pa_r  A \)^2  + \non
+&&  (D-3)\,\[ \( \pa_z C \)^2 -\(\pa_r  C \)^2 \]- 2\, \pa_z A \pa_z B +2 \,\pa_r A \pa_r B - \non
-&& 2\,(D-3)\,\(\pa_zB\pa_zC-\pa_rB\pa_rC\)  - {D-4 \over \rho\, \sqrt{f}} \( \pa_rA -\pa_rB \) + \non
 +&& {\sqrt{f} \over \rho} \left[ (D-4)\pa_r A +(D-2) \pa_rB -2\,(D-3)\pa_rC\right]=0.
\eea
As first observed by Wiseman \cite{Wiseman_RS},  the Bianchi
identities, $G^b_{a;b}=0$,  imply that the constraints satisfy
Cauchy-Riemann relations,
\bea
\label{cauchy_riemann}
 &&\pa_z \cU +\pa_r \cV = 0, ~~~~~ \pa_r \cU -\pa_z \cV =0, \non
&& \cU \equiv \sqrt{-g} G_r^z, ~~~~~ \cV \equiv \half \sqrt{-g} \(G_r^r- G_z^z \).
\eea
This in turn means that each one of $\cU$ and $\cV$ separately
satisfy the Laplace equation. Hence, the ``constraint rule''
follows: provided one of the constraints vanishes along the boundary
of the domain and the second one is zero at a single point of the
boundary---both constraints are guaranteed to vanish inside the
domain \cite{Wiseman_RS}.

The elliptic equations (\ref{Eqs}) are solved outside of the
horizon. The periodicity in $z$ and the reflection\footnote{An
appearance of the GL zero-mode spontaneously breaks the
translational invariance along $z$. The remaining symmetry to the
translation by half a period implies that the  branch of NUBS
solutions that emerges from the GL-point has reflection symmetry
about $z=0$.} symmetry at $z=0$ suggest the sufficient domain of
integration $\{ (r,z): 0 \leq z \leq L/2, r \geq 0 \}$.

The equations are subject to boundary conditions (b.c.) The
reflection symmetry at $z=0$ and the periodicity at $z=L/2$ are
translated into
\be
\label{reflection_bc}
 \pa_z \Psi|_{z=0} = \pa_z  \Psi|_{z=L/2}  =0  ~~ {\rm for} ~~ \Psi=A,B,C.
\ee
A regularity of (\ref{Eqs}) at the horizon supplies two conditions
\be
\label{horizon_bc}
\pa_r A|_{r=0}  = \pa_r C|_{r=0} =0.
\ee
However, we need three b.c. for our three fields.
The missing condition is obtained from the ``constraint rule", following which we impose $\cU=0$ along all boundaries.
 The constraint is automatically  satisfied along $z=0$ and $z=L/2$ due to (\ref{reflection_bc}), and  it vanishes
 exponentially at large $r$ (as does any z-dependence in a Kaluza-Klein background).
A regularity of  $\cU$ at the horizon yields
\be
\label{Bz}
\( \pa_z B -\pa_z A\)|_{r=0} =0,
\ee
and implies that the surface gravity is constant along  static horizon.
Integrating (\ref{Bz})  from $z=L/2$ we obtain a
Direchlet boundary condition for $B$,
\be
\label{B_hor}
B(0,z)=B|_{z=L/2} + A(0,z)-A|_{z=L/2}.
\ee
The integration constant $B_0\equiv B|_{z=L/2}$ is freely
specifiable; once it is chosen the NUBS solution is unique   for a
given $L$. $B_0=0$ corresponds to a unform string, a positive $B_0$
yields certain non-uniform string solution.\footnote{ An alternative
way to specify the integration  constant is to fix the surface
gravity $\kappa$. This determines B along the horizon as
$B=A-\log\kappa$. We checked that both approaches produce comparable
results. In either case, there is only one constant (either $B_0$ or
$\kappa$) that has to be chosen. We fix $B_0$ and
in this case the thermodynamical $\kappa$ becomes a derived
quantity.  }

Asymptotically we require that the spacetime is cylindrically-flat, $A=B=C=0$.
The leading fall-off of the fields in this region  is \cite{cagedBHsI}
\bea
\label{asymptotics}
&& A\simeq {a\over r^{D-4}}; ~~~~ B\simeq  {b\over r^{D-4}};  \non
&&  C  \simeq {c \log (r)  \over r} ~~ {\rm for} ~ D=5, ~~~~~    C  \simeq {c \over r}  ~~ {\rm for} ~ D>5 .
\eea
This form is useful in numerical implementations
whenever the asymptotic boundary is placed at a finite $r$.

Following the constraint rule and imposing $\cV=0$ at the asymptotic
boundary will relate the asymptotic constants\footnote{To this end,
one must also include the subleading term in the expansion of $C$,
that for $D>5$ is $\tilde{c}/r^2$.} in (\ref{asymptotics}). However,
in our numerical method which has the outer boundary at a finite
$r_a$,  we could get better overall accuracy  by simply enforcing
the fall-offs (\ref{asymptotics}) along $r_a$.  The constraints are
verified in appendix \ref{appendix_numeric}.

\subsection{Charges and geometry}
\label{sec_analysis}
Thermodynamic properties of the solutions are determined by the
asymptotic charges -- the mass and the tension, and by the horizon
measurables -- the surface gravity and the horizon area. The
variables are related by the generalized Smarr's formula. The
geometry of the black-strings is conveniently visualized by
embedding their horizon into flat space and by examining the proper
length of the compact circle along  $r=const$ slices.

\sbsection{Asymptotics and  charges}

The  charges are related to the asymptotic constants (\ref{asymptotics})  by \cite{cagedBHsI,HO1}
 \be
  \left[ \begin{array}{c} m \\ \tau\, L \\  \end{array} \right]= {\Omega_{D-3} \over 8 \pi} \left[ \begin{array}{cc}
  D-3 & -1 \\
  1   & -(D-3) \\
   \end{array} \right] \,
 \left[ \begin{array}{c} a+\half \rho_0^{D-4} \\ b \\ \end{array} \right]  \label{asymp_to_charges} \ee
where $  \Omega_{D-1} \equiv  D\,  \pi^{D/2}/\Gamma(D/2+1)$ is the surface area of a unit $\mathbb{S}^{D-1}$ sphere,
and  $\rho_0^{D-4}/2$ originates from the background $f(\rho(r))$, see (\ref{metric}). (We  set below $\rho_0=1$).
For convenience we use the ``asymptotic adapted" units in which the d-dimensional Newton's constant $G_d=G_D/L=1$.

The dimensionless charges, defined with respect to the critical uniform
black string's values (which can be read from (\ref{asymp_to_charges}) after setting $a=b=0$), are
\bea
\label{relative_charges}
\hm \equiv m/m_{c} &=& 1+ 2\, a -2\, b / (D-3) , \non
\htau  \equiv \tau/\tau_{c}&=& 1+ 2\, a -2\,b\,(D-3).
\eea
Notice the factor $D-3$ that in higher dimensions increases the significance of $b$ in the tension
computation and suppresses it relatively to $a$ in the mass calculation.

One can also define the relative tension, $n \equiv \tau L /m$.
Normalizing with respect to  $n_{c} =1/(D-3)$ one has
\be
\label{relative_n}
\hn \equiv  n/n_{c} =  \frac{1+ {2\, a }-{2\,b\,(D-3)}}{(D-3)\,(1+ 2\, a  )-2\, b }
\ee

\sbsection{Horizon variables}

The normalized temperature (or surface gravity) of the horizon reads
\be
\label{kappa}
\hT = \hkappa \equiv \kappa/\kappa_{c} = e^{A-B}|_{r=0},
\ee
where $T\equiv \kappa /(2\pi)$ and $\kappa_{c}=\half f'(\rho_0)=(D-4)/2$.
The normalized horizon area is given by
\be
\label{area}
\hA\equiv A/A_{c} = {1\over L} \int_0^L e^{B+(D-3)C}|_{r=0} \, dz,
\ee
where $ A_{c}=L\,  \Omega_{D-3} $ is the surface area of the critical string.

The horizon measurables are related to the asymptotic charges by
the generalized Smarr's formula \cite{cagedBHsI,HO1},
\be
\label{smarr}
(D-3)\,m = (D-2)\, S \, T   +\tau\, L ,
\ee
where $S \equiv A/4\,G_D$  is the Bekenstein-Hawking entropy.

The first law of thermodynamics relates variations of the thermodynamic variables.
If the asymptotic length of the compact circle is kept fixed (as in our case),
the first law reduces to
\be
\label{1stlaw}
dm = T dS ~~\Rightarrow ~~ d\hm =\hT d\hS \,{T_{c} \,S_{c} \over m_{c} }~~\Rightarrow ~~ d\hm =\hT d\hS \,{ D-4\over D-3}.
\ee

\sbsection{Geometry}

We examine
the black string's geometry by embedding the $r=const$ hypersurfaces into flat space,
\bea
\label{embedding}
ds_{flat}^2 &=& dZ_e^2 +R_e^2 d\Omega_{D-3}^2,\non
dZ_e&=& e^{B(r,z)}\,dz, \non
R_e&=& e^{C(r,z)} \rho(r).
\eea
The dimensionless proper length of the compact circle along an $r=const$ hypersurface is given by
\be
\label{Lprop}
\hL_{prop}(r)\equiv {1\over L} \int dZ_e(r) = {1\over L }\int_0^L e^{B(r,z)}\,dz.
\ee

In particular, the dimensionless proper length of the horizon is
$\hL_{hor} = \hL_{prop}(0)$.  The horizon's areal radius, $R_e(z)
=e^{C(0,z)}$, is not constant along the $z$-direction.

\subsection{Near the merger point}
\label{sec_nearmerger}

For progressively non-uniform strings the minimal areal radius of
the horizon, $R_{min}$, is steadily shrinking. Kol \cite{TopChange}
argues that in the regime when $R_{min}\rightarrow 0$  the local
geometry in vicinity of the ``waist" becomes cone-like (see also
additional arguments in  recent \cite{AKS}.) Moreover, there is the
power-low scaling (\ref{scaling}), which in $D<10$ is dressed with
periodic ``wiggles"  \cite{Kol_on_Choptuik}.

A direct comparison of the 6D solutions of \cite{Wiseman1} with the
cone metric was already carried in \cite{KolWiseman}, where in spite of low 
numerical resolution authors indeed found some evidence in favor of the cone geometry. In this paper,
by performing similar comparison, we provide additional, more
quantitative support to the cone conjecture. In addition, we test
the scaling (\ref{scaling}) and observe the power-law, but no
``wiggles'' in any dimensions.

It is convenient to reproduce here the relevant formulae from
\cite{TopChange,KolWiseman}  adopting them to our coordinates
(\ref{metric}). The comparison of the string's spacetime with the
cone is done in Euclidean signature. The transformation is achieved
by defining $t_E \equiv i \,t$ which takes the metric (\ref{metric})
to
\be
\label{metric_E}
ds^2=  f(\rho(r)) e^{2A} d t_E^2 +e^{2 \,B} (dr^2+dz^2) +\rho(r)^2 e^{2C} d\Omega_{D-3}^2 .
\ee
As usual in the Euclidean signature, the imaginary time  must be identified 
with period equal to the inverse temperature, $t_E \in [0,4\pi/(D-4) e^{A-B}]$.

\sbsection{The cone}

The Euclidean double-cone metric over the base $S^2 \times S^{D-3}$ reads
\be
\label{cone_metric}
ds^2_{cone}= d R^2 +R^2 \( {1\over D-2}\(d\chi^2 +\sin^2 \chi \kappa^2 d\tau^2\)+ {D-4 \over D-2} d\Omega_{D-3}^2 \)
\ee
where $\chi$ is the angle of $S_{\tau,\chi}^2$ ranging within $(0,\pi)$; $\tau=t_E$ and
therefore $\kappa=\half( D-4)e^{A-B}$ is the surface gravity of the black string.

Comparing (\ref{metric_E}) with (\ref{cone_metric}) one gets
\bea
\label{R_chi}
R(r,z)&=& e^{C(r,z)} \rho(r) \sqrt{D-4 \over D-2}, \non
\sin\chi(r,z)&=& { \sqrt{(D-4)\, f(\rho(r)) } \over \kappa \rho(r) } e^{A(r,z)-C(r,z)}.
\eea
Next, one rewrites the remaining
components of the cone metric in the $(r,z)$ coordinates,
\be
dR(r,z)^2+ {1\over D-2} R(r,z)^2 d\chi(r,z)^2 = c1 \,dr^2+c2\, dz^2+2\, c3 \,drdz,
\ee
where
\bea
\label{cone_components}
c1&=& (\pa_r R)^2 +{1 \over D-2} R^2 (\pa_r \chi)^2, \non
c2&=& (\pa_z R)^2 +{1 \over D-2} R^2 (\pa_z \chi)^2, \non
c3&=& (\pa_r R)(\pa_z R) +{1 \over D-2} R^2 (\pa_r \chi) (\pa_z \chi).
\eea
Comparing with (\ref{metric_E}) one concludes that the cone geometry will indeed approximate the
regions of the black string's spacetime  where  $c1=c2=e^{2\,(B,r,z)}$ and $c3=0$.

As an additional test, one can compare the Kretschmann scalar curvatures
of the numerical and the cone metrics.
For the cone this scalar is given by
\be
\label{Kre}
K \equiv R_{a b c d} R^{a b c d} \propto  {1 \over R^4},
\ee
where the proportionality constant depends on the dimension (in 5D
it is $48$, in 6D it is $72$, in 7D it is $320/3$ etc). For the
metric (\ref{metric_E}) the expression for $K$ is cumbersome but it
is straightforward to obtain.

\sbsection{The scaling and the wiggles}

In order to test the scaling (\ref{scaling}) one examines the
behavior of a characteristic length scale of the nearly pinching
strings as a function of some parameter along the NUBS branch. The
characteristic variable  whose behavior we test is  $R_{min}$.
Identifying a good parametrization\footnote{A good parameter should
be  defined on both sides of the merger, i.e. along BHs and NUBSs
branches, and it should be analytic at the merger point. In this
case, the scaling exponent would not be affected by a transformation
to another parametrization, provided the transformation is analytic.
In the critical collapse context this is related with the notion of
``universality'', i.e. independence of parametrization.}  appears to
be a difficult task in our case, since many potential
parameters, such as $\hat{\kappa}$ or $\hS$, for instance, are too
noisy at the near merger limit due to numerical errors. In this
paper, we use two different parameterizations: (i) By $p_1 \equiv
\hL_{hor}$, which monotonically grows and seems to tend to a finite
value, $\hL_{hor}^{max}$, at large nonuniformity; (ii) By $p_2\equiv
B_0^{-1}$, which is our low-level parameter, that apparently tends
to zero at the merger.  While the first parametrization is
continuous across the merger point\footnote{On the BHs side we can
define ``$\hL_{hor}$'' as a (normalized) proper length of the
compact circle along the symmetry axis $r=0$ and between the poles
of black hole along the horizon.}, it is not obvious  that it is
also analytic at this point. As for the second parametrization -- it
becomes not analytic at the merger. The scaling (\ref{scaling}) will
be confirmed if the graph of $\log(R_{min})$ against $\log(\delta
p_i)$ for $i=1$ and $i=2$ will be on average linear with periodic
wiggles about it.  The slope of the linear fit to the curve will
provide $\gamma_i$ and the period of the wiggles about the fit is
directly measured.

\section{Numerical implementation}
\label{sec_numerics}

We fix the asymptotic length of the compact circle to the critical
value $L=L_c=2 \pi/k_c$, where $k_c$ is the critical
Gregory-Laflamme wavenumber (see  table 1 in \cite{KolSorkin}), and
generate the NUBS solutions by varying the constant  $B_0\equiv
B|_{z=L/2}$ in (\ref{B_hor}).

The actual solution of the
elliptic equations (\ref{Eqs}) is obtained using \emph{relaxation}.
To this end the equations are discretized on a lattice covering the domain of
integration $\{ (r,z): 0 \leq z \leq L/2, 0 \leq r \leq r_a \}$, where $r_a$ is the
location of the outer boundary. The equations are written using a finite difference
approximation (FDA) which is second order in the grid spacings.
The relaxation of the non-linear equations (\ref{Eqs})
is achieved by incorporating a Newton iteration. Specifically, the
fields at an interior grid point $(i,j)$  are updated according to
\be
\label{relaxation_update}
\Psi^{new}(i,j)=\Psi^{old}(i,j)- \omega \[{ {\cal E}_\Psi(i,j) \over \pa {\cal E}_\Psi/\pa\Psi(i,j)} \]^{old},
\ee
for $\Psi=A,B$ and $C$; ${\cal E}_\Psi(i,j)$ is the FDA equation of
motion for $\Psi$ and $\omega$ is a numerical factor. In the basic
Gauss-Seidel relaxation method $\omega$ is set to unity. In our case
we find that choosing $\omega=1$ for all fields causes  divergence
of the code. However, taking $\omega_B <1$, that is under-relaxing
$B$, stabilized the scheme (usually we used $\omega_B \simeq 0.5$).

At the boundaries of the domain, boundary conditions (b.c.) are
incorporated. While, for Neumann b.c. the update procedure is
similar to (\ref{relaxation_update}), for Direchlet or Robin (mixed)
b.c. the equations are not solved. We use Direchlet condition
(\ref{B_hor}) to update $B$ at the horizon. The Robin boundary
conditions are used at the outer boundary, where $A$ and $B$ are
updated according to (\ref{asymptotics}): $A(z,r_a)= A(z,r_a-\Delta
r) [r_a/(r_a-\Delta r)]^{D-4}$ and
 $ B(z,r_a)=  B(z,r_a-\Delta r)[r_a/(r_a-\Delta r)]^{D-4}$, where
 $\Delta r $ is the grid-spacing. $C$ is updated according to
 \be
 \label{C_asymp}
 C^{new}(r_a) = (1-\mu_C) \,C^{old}(r_a) + \mu_C\, \[ C(z,r_a-\Delta r)\, {r_a\over r_a-\Delta r}  \]^{old}.
\ee
where $0\leq \mu_C < 1$ is the ``inertia" parameter. We generate most of our solutions
using large inertia i.e. $\mu_C \leq 0.05$. Using no inertia caused divergence. All other b.c. are Neumann.

The relaxation begins with specifying an initial guess for the
fields. Then, we sweep the grid using the red-black ordering and
update the fields according to (\ref{relaxation_update}). The sweeps
are repeated until the residuals drop below certain predefined
tolerance (typically, $10^{-7}$).

\subsection*{The hierarchy of meshes}

One of the disadvantages of relaxation methods is their relatively
slow convergence. A well-known technique intended to accelerate the
convergence is the full adaptive storage multi-grid method, see e.g.
\cite{ChoptuikUnruh}. Unfortunately, in our case we could not find
convergent multi-grid implementation\footnote{It doesn't blow up,
but the residuals just stop decreasing  below certain limit. See
appendix \ref{appendix_numeric} for further discussion of the
phenomenon.}.
\begin{figure}
\centering
\noindent
\includegraphics[width=11cm]{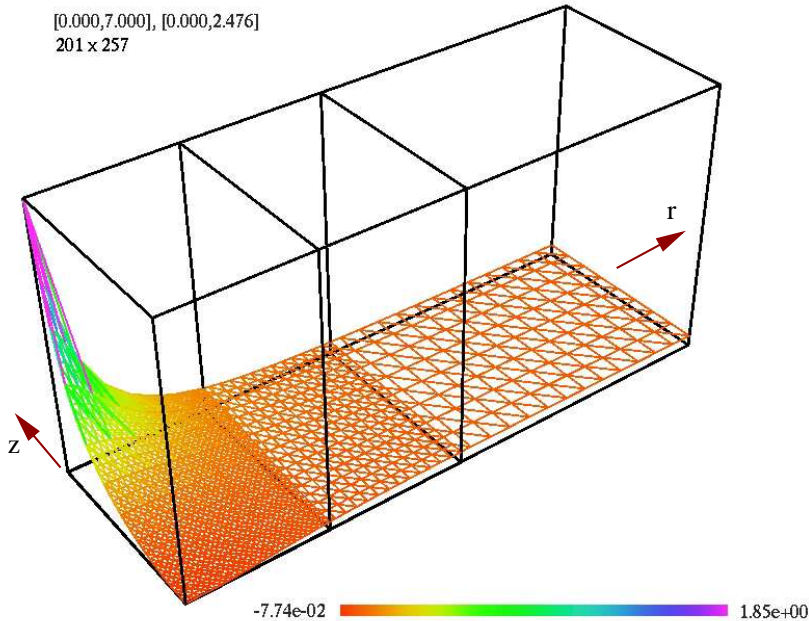}
\caption[]{The grid hierarchy, 3 meshes are shown. While the
coarsest mesh covers the entire domain, the finer grids extend only
to half size of the previous coarser mesh. The displayed field $B$
(as well as $A$ and $C$) is most variable near the horizon,
justifying this ``AMR''-type construction.} \label{fig_meshes}
\end{figure}

The method that we found to work well makes use of several grids but
in a different from the full multi-grid fashion.  We begin with some
initial guess on the coarsest grid and completely relax the
equations using the algorithm described in the previous subsection.
The solution is then interpolated into a finer grid where it is
regarded as an initial guess. The equations are relaxed on this grid
too, and the solution is passed to the next finer grid. The
procedure is repeated on a desired amount of grids. The one-way
manner that we propagate the solution along the grid hierarchy is to
be contrasted with the bi-directional (V-cycle) communication
between grids that is utilized by full multi-grid methods.

We construct the mesh hierarchy using $2:1$ refinement. Furthermore,
in this sequence the radial extension of a finer mesh is taken to be
half the extension of the previous mesh. The truncation is basically
done because the high resolution delivered by dense grids is only
necessary near the horizon where the fields are most
variable\footnote{ Hence our refinement is in tune with the adaptive
mesh refinement (AMR) approach.}. Note that number of the grid
points in radial direction is conserved in this hierarchy.
 The boundary conditions used on finer grids do not
change, except at the outer boundary, where we keep the fields at
the values obtained on the coarsest grid. We verify that  no
mismatch  between our NUBS solutions on any two subsequent grids
arises along this boundary. Figure \ref{fig_meshes} illustrates the
hierarchy and the field $B$ on it in 6D.

The initial guess on the coarsest mesh requires some caution. A
priori we don't have a natural choice, so we use the simplest
$A=B=C=0$ substitution. We found that this guess is relaxed
effectively only for $B_0$ below certain value (which roughly
corresponds to mildly deformed strings) for larger $B_0$'s it leads
to a very slow convergence or to divergence. However, using a
solution obtained at smaller $B_0$ as an initial guess for higher
$B_0$'s prevented the code from crushing and  improved the
convergence rate.

Despite using this trick, on each particular initial grid we were
able to find solutions only up to certain maximal $B_0$. For larger
$B_0$'s the method diverges no matter how we tune up the parameters.
In this case, we could continue to larger $B_0$'s by increasing the
density of the initial mesh. A significant gain in the convergence
rate was achieved by feeding the available lower resolution solution
as an initial guess.

As a result,  after finding the NUBS branch we end up with a family
of several initial meshes having different resolutions, and the
multi-grid hierarchy attached to each one of them. We verify that
the physical measurables obtained on various meshes within the
hierarchy converge quadratically. When we had a family of several
initial grids with overlapping parameters we verified the
convergence rate as a function of the resolution on the initial
mesh. Figures \ref{fig_S_6D},\ref{fig_T_6D} and \ref{fig_mass6D} are
made in six dimensions where we have such an overlap. The
convergence rate of the quantities shown in figures is nearly second
order.

A verification of the constraints as well as further technical details of
the numerical procedure are found in appendix \ref{appendix_numeric}.

Overall, we find that the method performs very efficiently in lower
dimensions and small nonuniformities. For large nonuniformities the
convergence slows down and, in general,  violation of the
constraints grows.  We could overcome the problems and accurately
get the solutions with stronger horizon deformations by increasing
the resolution of the numerical lattice. We had to stop when the
mesh-size made the relaxation time unreasonably long.  In higher
dimensions the problems are more pronounced and begin earlier. In
addition, the range of the parameters $\omega_B$ and $\mu_c$, for
which the relaxation converges, shrinks. In practice, using our
current method we were only able to get the NUBS solutions in the
dimensions between six to eleven. The following section summarizes
our findings.

\section{Results }
\label{sec_results}

We construct the branch of  non-uniform black strings in dimensions
$6 \leq D \leq 11$ and extend it deep into non-linear regime.  In 6D
we find good agreement with previous results
\cite{Wiseman1,Oldenburg_BS} in the range of $\lam$'s obtained in
these works, $\lam \simeq 4$ in  \cite{Wiseman1}, and $\lam \simeq 6
$ in  \cite{Oldenburg_BS},  and continue the branch up to $\lam
\simeq 8$. Below we describe the behavior of the characteristic
physical variables  and  their dependence on the dimension.

\begin{figure}
\centering
\noindent
\includegraphics[width=10cm]{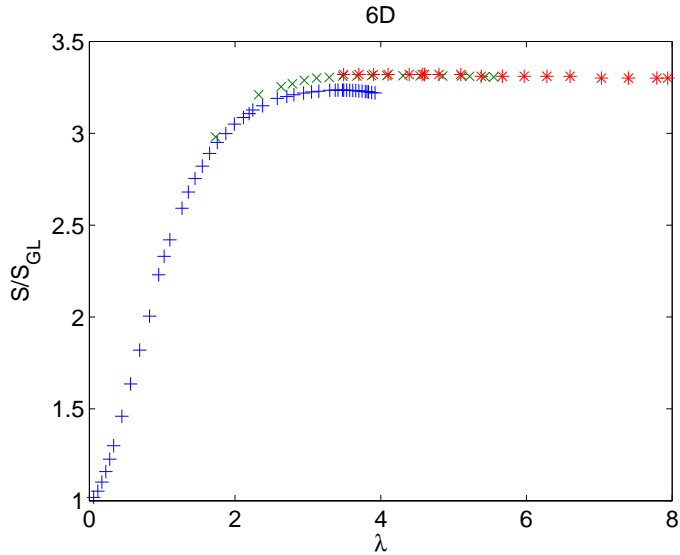}
\caption[]{The normalized entropy (\ref{area}) of NUBSs in 6D as a
function of $\lam$. Shown are the solutions obtained using three
mesh resolutions: low (pluses), medium (crosses) and high (stars),
obtained by halving the grid-spacings. In the region where all three
overlap we observe nearly second order convergence. The  NUBSs are
more entropic than that of the critical uniform string. In the limit
of large $\lam$ the entropy reaches saturation. The bending that
shows up beyond certain $\lam$ at lower resolution, at a higher
resolution moves to larger $\lam$'s  in the manner
 that doesn't form clearly convergent sequence. }
\label{fig_S_6D}
\end{figure}

The typical behavior of the (normalized) entropy (\ref{area}) in the
verified range of the dimensions is depicted in figure
\ref{fig_S_6D}. The key feature here is that the relative entropy
grows with $\lam$, so that  NUBSs are more entropic than the uniform
critical string. In $D\leq 9$ we find that at large $\lambda$'s the
entropy asymptote to certain limiting value. These limiting values
are listed in table \ref{table_variables} and they estimate the
entropy of the merger string. In higher dimensions our numerical
method does not allow extending the branch that far,  in this case
the numbers in the table are only lower bounds on the asymptotic
$\hS$'s. We find that the $\hS(\lam)$ curve steepens as $D$ grows
and the saturations occurs earlier. For example, while in 6D this
happens around $\lam \simeq 3$, in 9D $\hat {S}$ reaches
saturation just above $\lam \simeq 1.5$. The entropy of the nearly
pinching strings decreases with the dimension.

We note that the $\hat {S}$ curve has the back-bending reported in
\cite{Oldenburg_BS}. However, the variation of $\hS$ in the bending,
$\delta S/S$, is less than a percent in our range of $\lam$'s. In
addition, from the figure we learn that the specific value of
$\lam_{bb}$ beyond which the curve bends depends on the resolution
of the numerical lattice used to find the solution, in a way that an
increase in resolution increases $\lam_{bb} $. In 6D (shown in
figure) we use a sequence of three resolutions with grid spacings
$h$, $h/2$ and $h/4$. The resulting $\lam_{bb}$'s are approximately
$3.5, 4.5$ and  $5.5$, which doesn't seem to form a convergent
sequence. Similar behavior is gotten in higher $D$'s as well.

\begin{figure}
\centering
\noindent
\includegraphics[width=10cm]{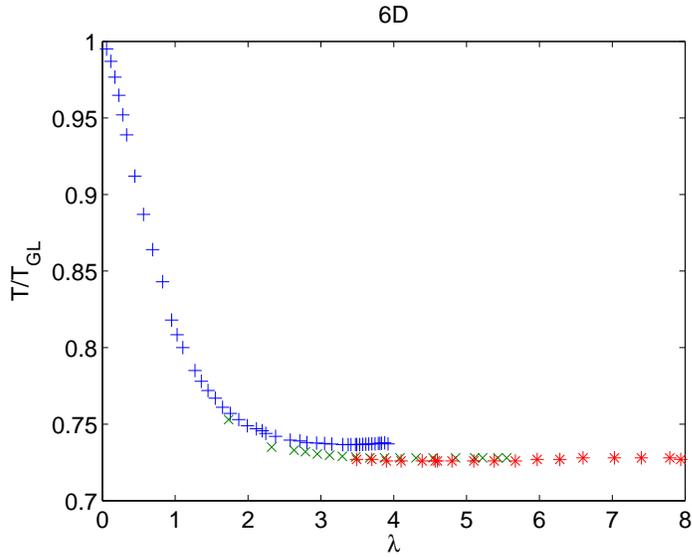}
\caption[]{The temperature in 6D as a function of $\lam$.
Same conventions as in figure \ref{fig_S_6D}. }
\label{fig_T_6D}
\end{figure}

The typical dependence of the temperature (\ref{kappa}) on $\lam$ is
depicted in figure \ref{fig_T_6D} which shows that the NUBSs are
cooler than the critical GL string. The temperature  decreases with
$\lam$ and in $6\leq D \leq 9$ it is seen to tend to certain
limiting values. In higher dimensions our numerics loses stability
before the saturation is achieved. The minimal  attained
temperatures are listed in table \ref{table_variables}. The features
such as the steepening of the $\hT(\lam)$ curve in higher dimensions
and the bending are similar to those observed for the entropy. The
temperature of the nearly pinching strings in various dimensions
increases with $D$.

\begin{figure}
\centering
\noindent
\includegraphics[width=10cm]{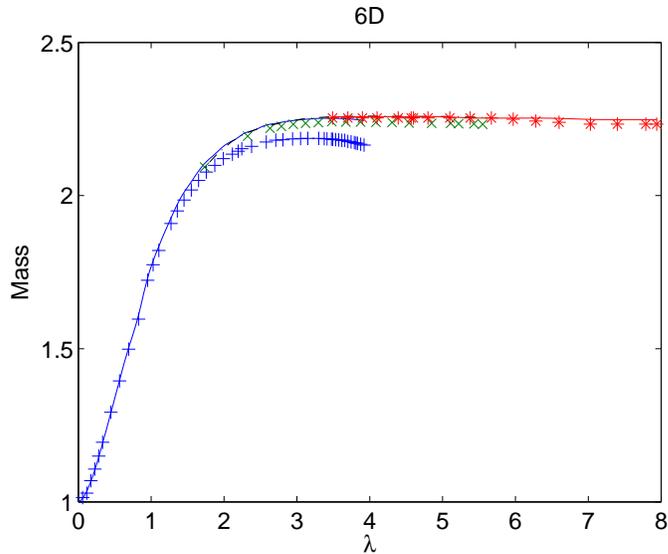}
\caption[]{The mass as a function of $\lam$ in 6D, same conventions
as in figure \ref{fig_S_6D}. The markers designate the mass computed
from (\ref{relative_charges}), the solid lines indicate the result
obtained by integration of the first law, see (\ref{m_1law}). Both
methods agree well and the agreement improves at higher resolution.
} \label{fig_mass6D}
\end{figure}

In figure \ref{fig_mass6D} we plot the normalized mass as a function
of $\lam$. The mass is computed in two ways: (i) From the asymptotic
charges according to (\ref{relative_charges}), the result is
designated by markers; (ii) Integrating the first law (\ref{1stlaw})
according to (\ref{m_1law}), this is represented by solid line.  The
figure demonstrates that both methods yield comparable results and
the agreement further improves as we increase the resolution. In all
verified dimensions the mass along NUBS branch is higher than that
of the critical string. All other features of the mass curve are
similar to those of the entropy curve.  At large $\lambda$, in
$6\leq D\leq 9$, the mass reaches saturation, the asymptotic values
recorded in table \ref{table_variables}. In higher dimensions the
table shows only the maximal masses obtained rather than
asymptotics. The mass of the nearly pinched strings is a decreasing
function of the dimension.

\begin{figure}
\centering
\noindent
\includegraphics[width=10cm]{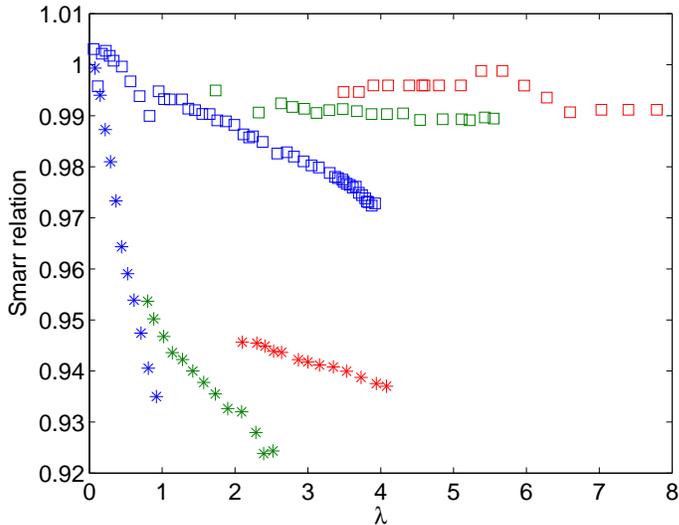}
\caption[]{The Smarr formula (\ref{smarr}). Shown are the ratios
between the two sides of (\ref{smarr}) in 6D (squares) and 8D
(stars).  In each dimension, there are three sequences of points 
obtained at three numerical resolutions, with the sequence at largest $\lam$ corresponding to the largest resolution. 
In either case, the points concentrate near $1$ such that the
violation of Smarr's relation never exceeds the $10 \%$ level.}
\label{fig_smarr}
\end{figure}

Smarr's formula (\ref{smarr}) relates the local horizon variables
$S$ and $T$ with the asymptotic ones, $m$ and $\tau$,  and as such
it is an important indicator of \emph{global} accuracy of a
numerical method\footnote{It turns out that the tension $\tau$ along
the NUBS branch cannot be extracted accurately enough in our
numerical implementation.  However, this fact does not have major
impact on the accuracy of the Smarr relation. See appendix
\ref{appendix_numeric} for analysis and discussion.}. In figure
\ref{fig_smarr}  we plot the ratio between the two sides of
(\ref{smarr}) in six and eight dimensions. We observe that in
general the formula is well satisfied. The scattering  of the points
does not exceed  $10\%$, a large fraction of which is probably due
to inaccurate  $\tau$.  For a given resolution the accuracy
decreases for progressively non-uniform solutions. However, the
accuracy improves at higher resolutions,  indicating convergence.

\begin{figure}
\centering
\noindent
\includegraphics[width=10cm]{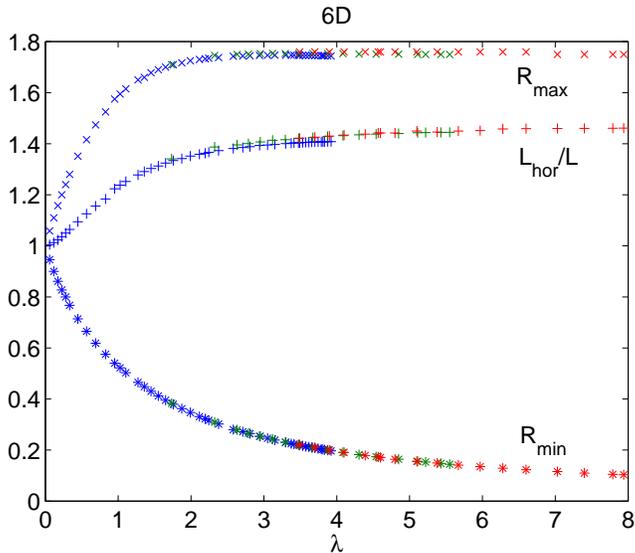}
\caption[]{The typical behavior of the geometric variables:
$R_{min}$ decreases steadily with $\lam$, $R_{max}$ and $\hat{L}_{hor}$
asymptote to finite values. Same conventions as in figure \ref{fig_S_6D}.  }
\label{fig_geometrical_vars}
\end{figure}

To analyze the geometry of the NUBSs we plot in figure
\ref{fig_geometrical_vars} the minimal and the maximal areal radii
of the horizon and its proper length along the compact circle as a
function of $\lam$.  We find that while $R_{min}$ shrinks
monotonically with $\lam$, $R_{max}$ and $\hL_{hor}$ asymptote to
finite values in the large $\lam$ limit. The behavior of the curve
$R_{max}(\lam)$ is again similar to the behavior of the  entropy.
The $\hL_{hor}(\lam)$ curve is similar too, with the only difference
that it does not have any back-bending. As the dimension grows the
asymptotic values of $R_{max}$ and $\hL_{hor}$ are getting smaller,
see table \ref{table_variables}.

Figure \ref{fig_cone-sketch} shows the embedding (\ref{embedding})
of the black string horizon into flat space. The maximal and the
minimal radii of the horizon are explicitly seen to occur at $z=0,L$
and at $z=L/2$ respectively.  The proper length of the circle along
$r=const$ is maximal at the horizon but for large $r$, $\hL
\rightarrow 1$ exponentially  fast. Note that the transition from a
uniform  to a non-uniform black string causes an expansion of the
compact circle near the horizon. In this sense we have an Archimedes
effect for nonuniform black strings. The effect weakens in higher
dimensions, see table \ref{table_variables}.

\begin{table}[h!]
\centering \noindent
\begin{tabular}{|c||c|c|c|c|c|c|c|c|c|}\hline\hline
$D$ & $\lambda_{max}$ & $\hm$  & $\hT $ &  $\hS$&   $R_{max}$ &  $R_{min}$ &  $\hat{L}_{hor}$   \\
\hline\hline
6 & 8 & 2.26 & 0.73  & 3.3 & 1.75& 0.103 & 1.45\\ \hline
7 & 4.45 & 2.37 & 0.793 & 3.25& 1.57& 0.16& 1.43\\ \hline
8 & 4.1  & 2.28 & 0.86  & 2.83& 1.42& 0.156& 1.34\\ \hline
9 & 1.7   & 1.98 & 0.91  & 2.3 & 1.32 & 0.3& 1.23\\ \hline
10 & 0.46& 1.5& 0.96 &1.7 & 1.22& 0.64& 1.07\\ \hline
11 & 0.2& 1.14& 0.99  &1.22 & 1.13& 0.81& 1.04\\ \hline
\hline
  \end{tabular}
\caption[]{The thermodynamic and the geometric variables at the
maximal $\lam$ to which we could extend the NUBS branch. In the
dimensions between six to nine the behavior of the variables is
similar to this plotted in figures \ref{fig_S_6D}, \ref{fig_T_6D},
\ref{fig_mass6D} and \ref{fig_geometrical_vars}. Namely, in these
dimensions the listed numbers are close to  the asymptotics.
($R_{min}$ does not reach saturation but continuously decreases. The
shown values correspond to the minimal attained $R_{min}$.) In
higher dimensions the shown numbers are not asymptotics, but only
bounds on  the true limiting values.
 }\label{table_variables}
\end{table}

\subsection*{The local cone}

\begin{figure}
\centering
\noindent
\includegraphics[width=12cm]{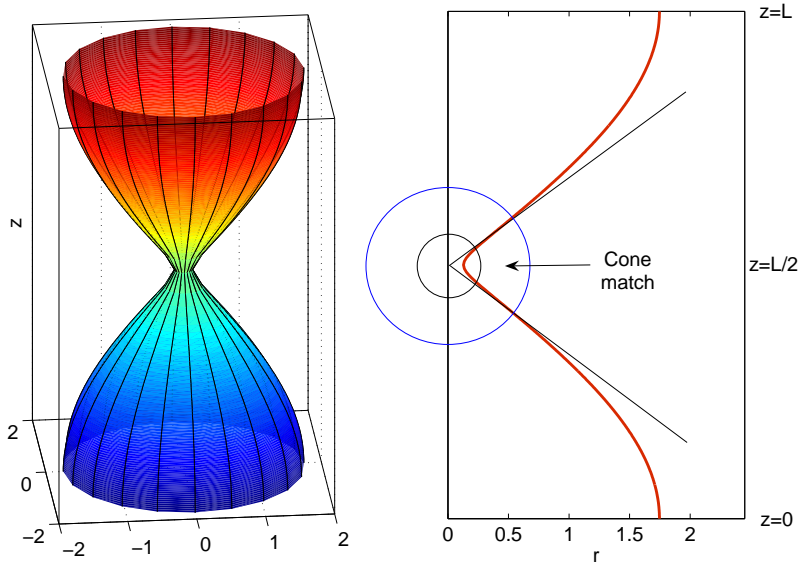}
\caption[]{The embedding of the string's horizon into flat space (\ref{embedding}).
The right panel shows
the cone and the region (between the circles) where the local geometry
is compared with the cone. }
\label{fig_cone-sketch}
\end{figure}

We compare the local geometry in vicinity of the ``waist'' of
strongly nonuniform strings with the cone metric. The relevant
spacetime region is illustrated in figure \ref{fig_cone-sketch}. The
cone sketched in the right panel is predicted to approximate the
local merger geometry of the marginally  pinching string. In this
case one would expect that a non-pinching, but strongly deformed
string still carries signs of the cone geometry in the region
bounded by the circles shown in figure. We will now demonstrate that
with high precision the geometry near the ``waist"  is indeed
cone-like.

\begin{figure}
\centering
\noindent
\includegraphics[width=10cm]{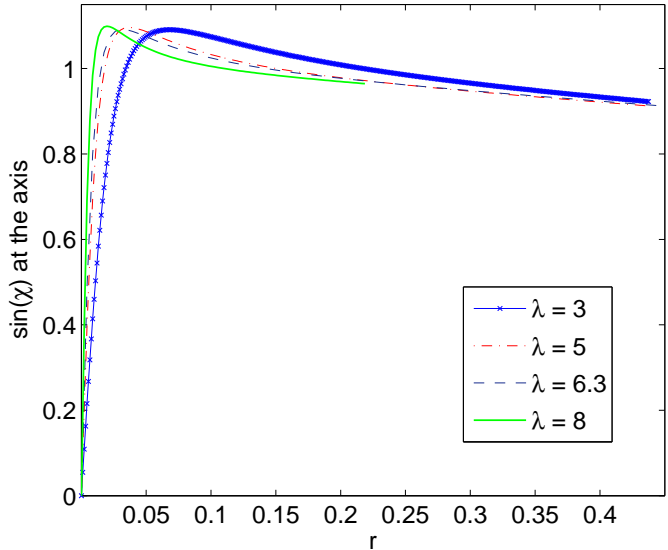}
\caption[]{$\sin\chi$ along the $z=L/2$ axis as determined from
(\ref{R_chi}) in 6D. For the cone $\sin\chi|_{axis}=1$, but for a
finite $\lam$ NUBS this is not so. Since the cone tip is missing,
the curve initially climbs steeply to values greater than $1$, stays
close to unity and finally deviates from one farther away from the
waist. For increasingly nonuniform strings the peak tends to the
origin and the entire curve gets consistently closer to unity. This
indicates that the cone tip is closer approached. The markers in one
of the graphs designate the actual resolution of the numerical
lattice. } \label{fig_sinchi}
\end{figure}

We explicitly compare the numeric and the cone metrics in six
dimensions where we have our most non-uniform solutions and the
highest resolutions. First, according to (\ref{R_chi}) we obtain $R$
and $\chi$ in terms of $(r,z)$. The extraction of $\chi$ appears to
be delicate \cite{KolWiseman} because $\chi$ is obtained from
$\sin\chi$, which for our numerical solutions with finite $\lam$
behaves somewhat differently from what it should for the cone.
Specifically, at the axis $z=L/2$, that corresponds to $\chi=\pi/2$,
the cone must have $\sin\chi=1$. However, since the cone apex is
missing at finite $\lam$, $\sin\chi$ obtained from (\ref{R_chi})
happens to be greater than $1$ at small $r$'s, see figures
\ref{fig_cone-sketch} and \ref{fig_sinchi}. At the same time, figure
\ref{fig_sinchi} shows that along the axis sine is close to $1$. The
peak in the curve, that appears because the cone's tip is missing,
drifts to the origin as $\lam$ grows indicating that the cone apex
becomes resolved better in the large $\lam$ limit.

Clearly, having $\sin\chi|_{axis}>1$ obstructs the extraction of
$\chi$. A practical remedy to the problem was suggested in
\cite{KolWiseman}, where the sine was ``renormalized'' by its value
at the axis, forcing it to the correct, $\sin\chi_{axis}=1$,  value.
The renormalization is achieved by
\be
\label{sinchi_corr}
\sin\chi(r,z)|_{corrected}={\sin\chi(r,z)\over \sin\chi_{axis}},
\ee
where in order to normalize outside of the axis one finds $\sin\chi$
at the axis and proceeds along $R=const$ curves using this value.

\begin{figure}
\centering
\noindent
\includegraphics[width=10cm]{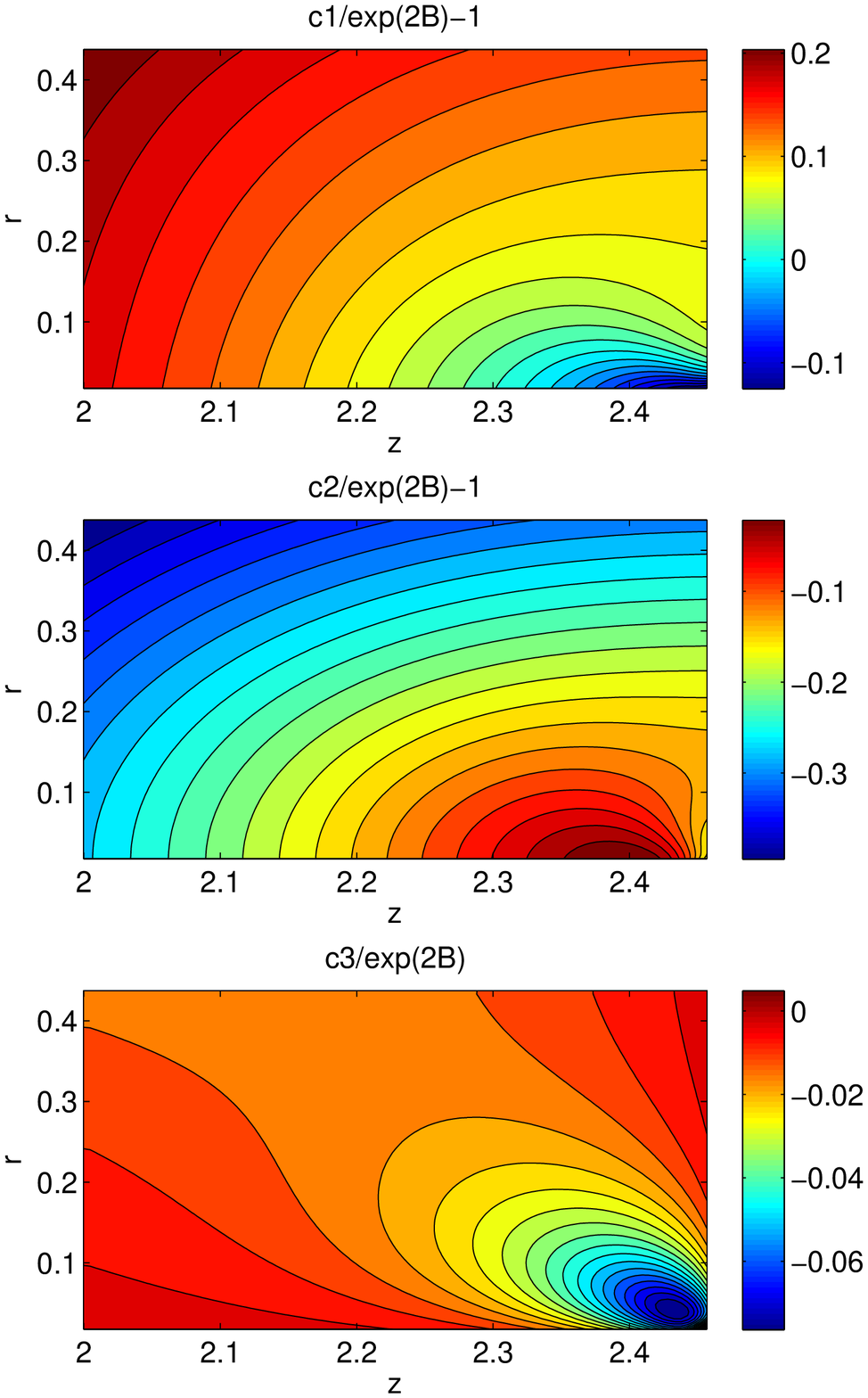}
\caption[]{6D; $\lam \simeq 6.3$. The metric components
(\ref{cone_components}) that must satisfy $c1=c2=e^{2\,B}$ and
$c3=0$ if the cone indeed approximates local  geometry near the
``waist" which is located at the bottom right corner. $c1$ and $c2$
agree very well with the cone prediction: we see approximately
$90\%$ match slightly away from the corner. $c3$ is two orders of
magnitude smaller than $c2$ and $c1$, that is consistent with the
cone prediction as well. The sub-domain shown here is covered with
the numerical lattice of approximately $200 \times 200 $
mesh-points.   } \label{fig_cone_comp}
\end{figure}

\begin{figure}
\centering
\noindent
\includegraphics[width=10cm]{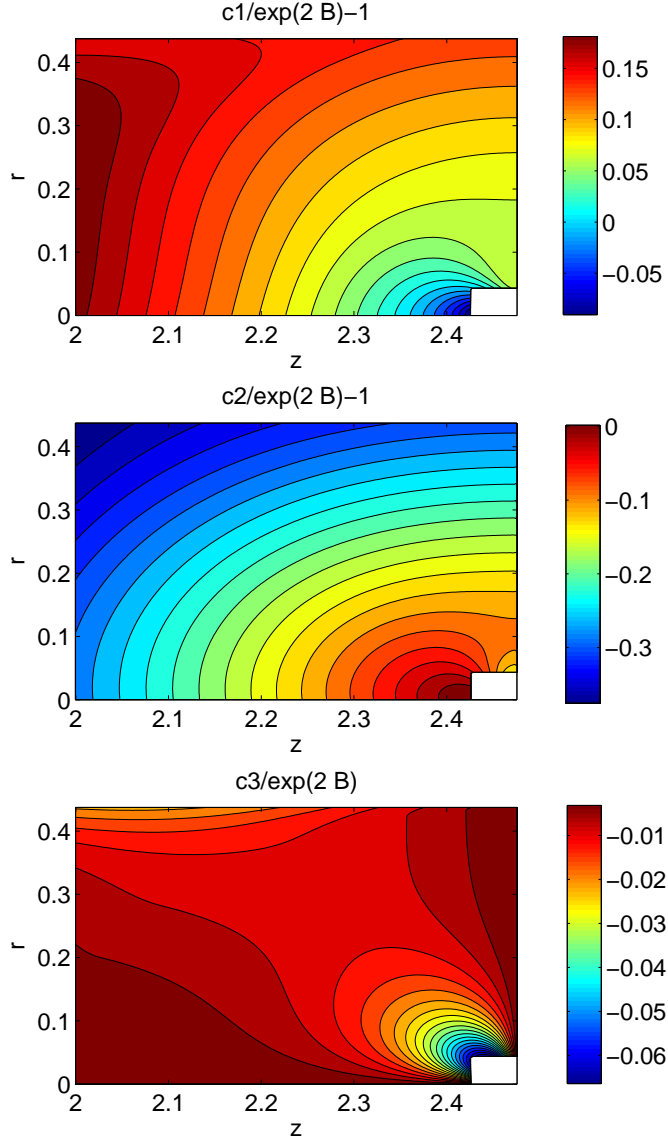}
\caption[]{6D. $\lam \simeq 8$ and we use same conventions as in
figure \ref{fig_cone_comp} but remove the apex. The agreement with
the cone prediction is somewhat better than in figure
\ref{fig_cone_comp}. } \label{fig_cone_comp1}
\end{figure}

Having obtained $R$ and $\chi$,  we determine the metric components
(\ref{cone_components}) $c1,c2$ and $c3$ and plot them in figures
\ref{fig_cone_comp} and \ref{fig_cone_comp1} for $\lam\simeq 6.3 $
and $\lam\simeq 8$ respectively. There is an excellent agreement
with the cone prediction. It is required that $c1=c2=e^{2\,B}$ and
from the figures we learn that while $c2$ and $c1$ differ from
$e^{2B}$ by several tens of percents far away from the waist, the
match grows to a level of more then $90\%$ closer to the corner.
Expectedly, the agreement worsens again in the immediate vicinity of
the apex, see figure \ref{fig_cone-sketch}. As for $c3$, which must
vanish for the cone -- it is two orders of magnitude smaller than
$c2,c1$ in the relevant region, and this is indeed consistent with
the prediction.
\begin{figure}
\centering
\noindent
\includegraphics[width=10cm]{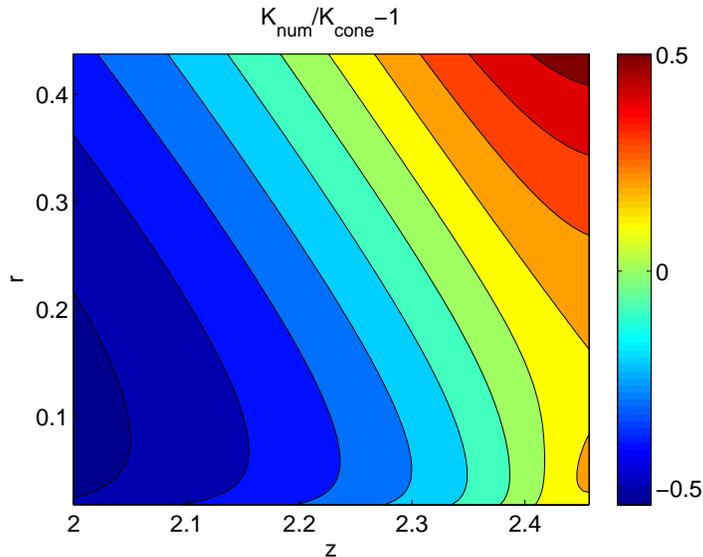}
\caption[]{6D. A comparison of the Kretschmann invariants computed
for the cone and the numerical black-string with $\lam \simeq 6.3$.
A good match is in the vicinity of the ``waist'', which is located
at the right bottom corner of the figure. } \label{fig_Kcompare}
\end{figure}
\begin{figure}[h!]
\centering
\noindent
\includegraphics[width=8cm]{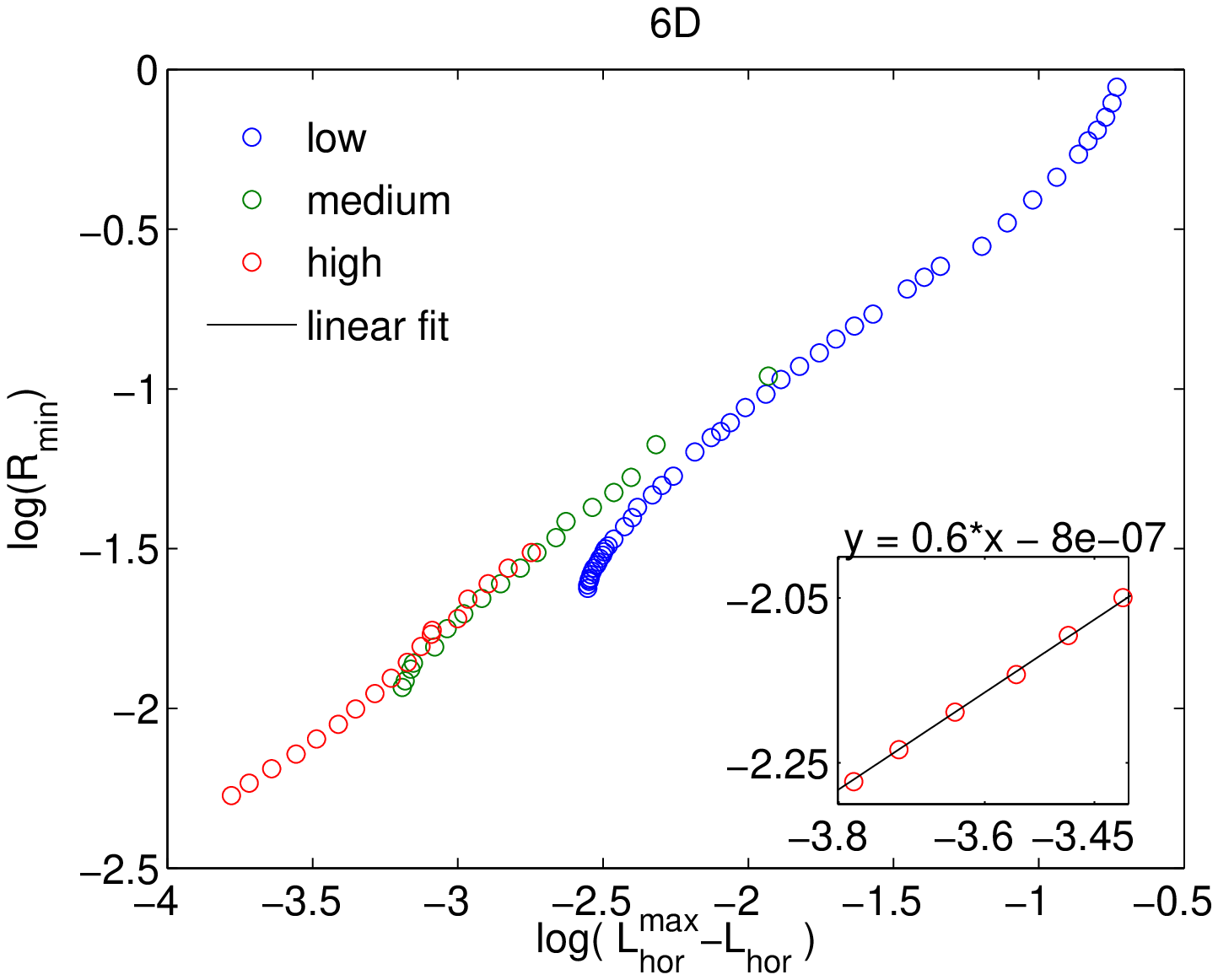}
\includegraphics[width=8cm]{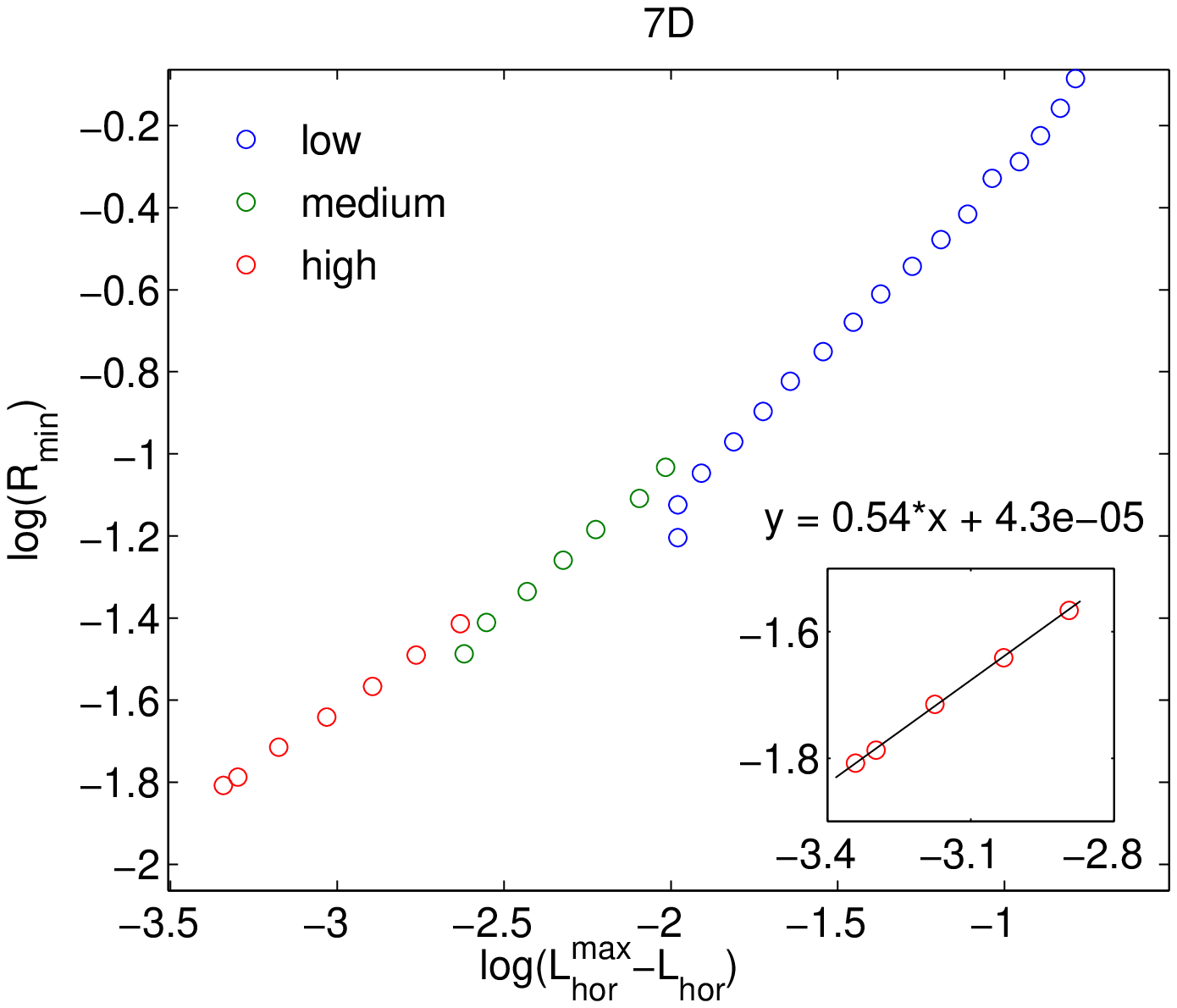}
\includegraphics[width=13cm]{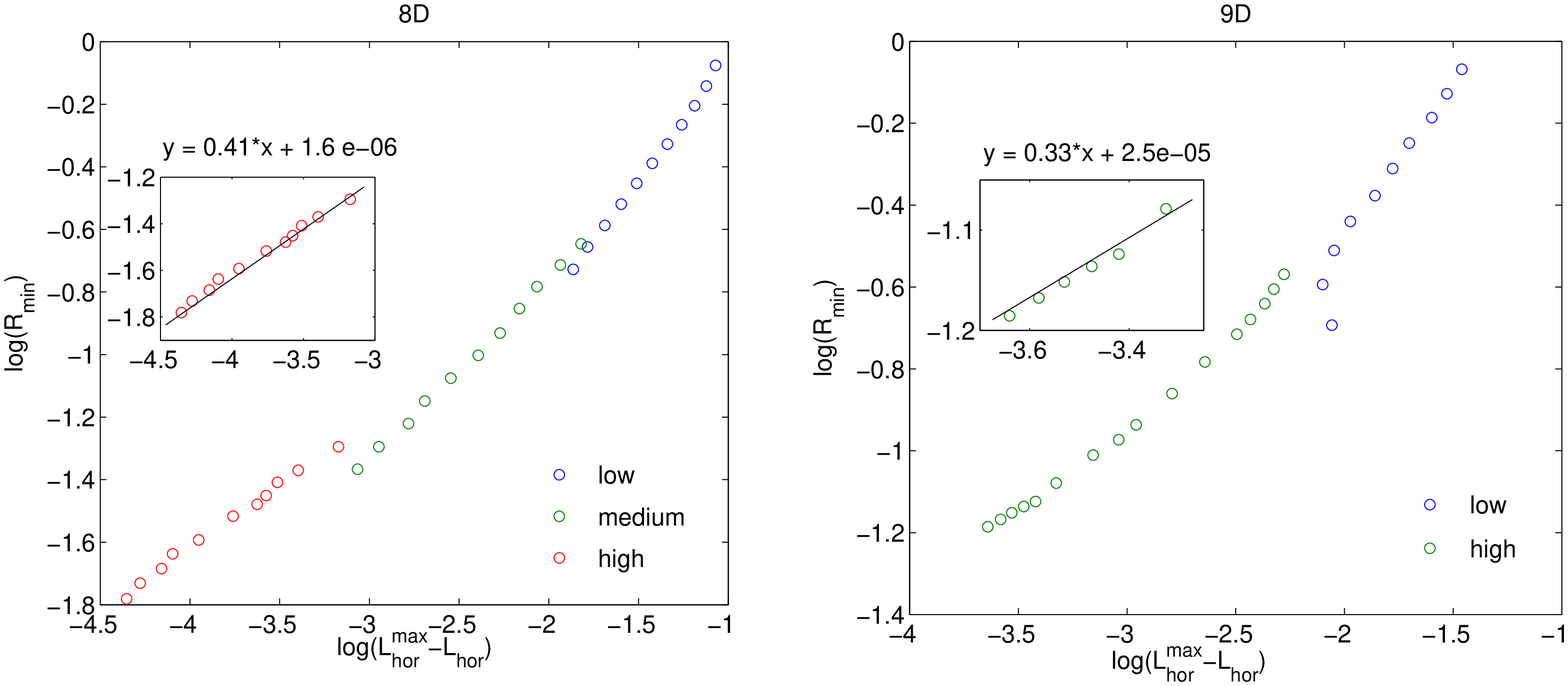}
\caption[]{The scaling of $\log(R_{min})$ as a function of $\delta p
= \log(\hL_{hor}^{max}-\hL_{hor})$ in various dimensions. 
We show the sequences of data points obtained at several resolutions, 
with the largest resolution sequence corresponding to the smallest $\delta p$.  The inset shows only the
highest resolution data that corresponds to the most deformed
strings. The solid line, which is the linear fit, is seen to
approximate the data points well. This indicates that at large
$\lambda$'s we have a power-low.  The corresponding exponents are
given by the pre-factors in the displayed fitting equations. }
\label{fig_Rmin_scaling}
\end{figure}
\begin{figure}[h!]
\centering
\noindent
\includegraphics[width=12cm]{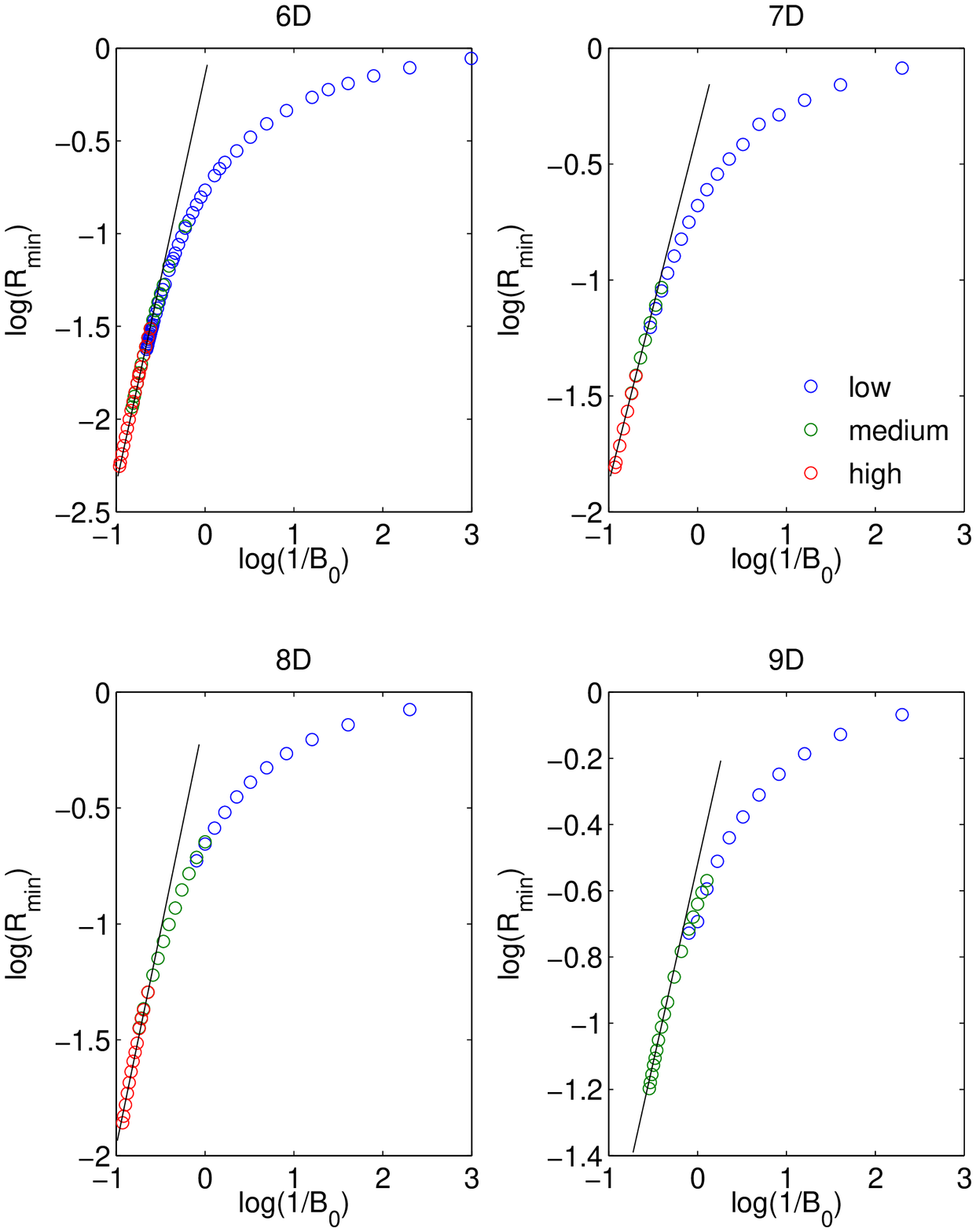}
\caption[]{The scaling of $\log(R_{min})$ as a function of $\delta p
= \log(B_0^{-1})$ in various dimensions. Same conventions as in figure \ref{fig_Rmin_scaling}.  In the limit of large deformations
the scaling is clearly a power-law as indicated by the shown linear
fit. } \label{fig_Rmin_scalingB0}
\end{figure}

Another  test involves comparison of the Kretschmann curvature
invariants computed for the numerical and the cone metrics, see
figure \ref{fig_Kcompare}. The plot indicates that the invariants
are comparable around the ``waist", where the mismatch between them
is only about $10 \%$.

Next we verify the scaling (\ref{scaling}). A characteristic
quantity with dimensions of length whose scaling we will test is the
minimal areal radius of the horizon $R_{min}$. It is measured
directly from the metric and hence it is the cleanest, from the
numerical point of view, variable. One of the parameterizations of
the NUBS branch that we use in this paper is given by $p \equiv
\hL_{hor}$. The horizon length apparently remains finite at the
pinch-off, and as we saw it doesn't exhibit any bending.  Assuming
that $\hL_{hor}$ is growing monotonically we obtain the limiting
length, $p*=\hL_{hor}^{max}$ by extrapolation.

Figure \ref{fig_Rmin_scaling} plots $\log(R_{min})$ as a function of
$\delta p \equiv \log(\hL_{hor}^{max}-\hL_{hor})$ in six, seven,
eight and nine dimensions. Figure shows that for small $p$'s the
scaling of $R_{min}$ is not a power-law. However, the behavior is
approaching a power-law dependence in the $p \rightarrow p*$ limit.
This is most pronounced in 6D where we have the longest data series.
The ``runaways" of the data points that appear at the end of each
sequence with a given resolution is probably related to the loss of
accuracy.

The insets show several points that correspond to the largest
non-uniformity. These points can be fit quite accurately by a
straight line (shown), and this indicates the power-law scaling. The
corresponding exponents are given by the slope of the fits and they
are summarized in table \ref{table_gammas}. These are only
approximate values because of the numerical errors related to the
fitting procedure (whose accuracy is  within $5\%$), and other
manipulation such as extrapolation that we used to find $p*$ etc.
Besides, the values in table should strictly speaking be regarded as
upper bounds on the exponents, rather than the exponents themselves.
It is hard to say how close to the actual $\gamma$'s the bounds are,
but at least in six and seven dimensions they seem to provide a good
estimate, see figure \ref{fig_Rmin_scaling}.
\begin{table}[h!]
\centering \noindent
\begin{tabular}{|c|c|c|c|c|}\hline
 D &  6 & 7  & 8  & 9\\ \hline
 $\gamma$ in $\hL_{hor}$ &0.6 & 0.54 & 0.41 & 0.33  \\  \hline
 $\gamma $ in $ B_0^{-1} $ &2.1 & 1.8 & 2.2 & 1.4  \\  \hline
  \end{tabular}
 \caption[]{The scaling exponents $\gamma$ in $R_{min} \propto \delta p^{\gamma}$ in various dimensions for two parameterizations.
The  first row shows the upper bounds on $\gamma$ in the
parametrization by $\hL_{hor}$. The second row is the lower bound on
$\gamma$ in the parametrization by $B_0^{-1}$. Distinct $\gamma$'s
for a given $D$ reflect the fact that the transformation between the
parameterizations is not analytic.} \label{table_gammas}
\end{table}

In principle, we could parameterize the NUBSs branch differently,
for instance,  by $\kappa$ or by $\hS$. In practice however, all
measurables except $\hL_{hor}$  have the back-bending that hampers
determination of their asymptotic values, which are crucial for
forming $\delta p$. On the other hand, the control parameter,  $
B_0^{-1}$, that we use to generate our solutions is a monotonic
function of $\lam$, and it naturally parameterizes the NUBSs branch.
A disadvantage of this parametrization is that it is not defined on
the BHs side. Figure \ref{fig_Rmin_scalingB0} depicts
$\log(R_{min})$ vs $\log B_0^{-1}$.  The behavior is not a power-low
for mildly deformed strings, however as indicated by the straight
lines shown in figure, the dependence becomes such in the
near-merger regime. The slopes of the lines provide lower bounds on
the critical exponents $\gamma$, we list these in table
\ref{table_gammas} and observe that $\gamma \sim 2$ in all recorded
dimensions. Comparing $\gamma$'s in both parametrization we see that
they are distinct, therefore the parameterizations are not
analytically related.

Having estimated the exponents we can ask about the fine structure,
or the "wiggles", that according to the prediction should appear as
a wavy structure about the linear fit in figures
\ref{fig_Rmin_scaling} and \ref{fig_Rmin_scalingB0}. By examining
figure \ref{fig_Rmin_scaling} one infers that in six and seven
dimensions there are no any wiggles in the range of the achieved
nonuniformity, and in higher dimensions the accuracy of the data is
insufficient for deciding about existance/absence of the wiggles.
From figure \ref{fig_Rmin_scalingB0} it follows that the wiggles do
not appear in any of the plots.

\section{Discussion}
\label{sec_discussion}

Our numerical construction of the branch of the  NUBS solutions
confirms the trends observed in the perturbative analysis of
\cite{CritDim} and lends further support to the phase diagram
\ref{fig_phasediagram}. We managed to extend the solutions branch
into strongly nonuniform regime and this allowed us to estimate the
physical parameters of the nearly pinching strings, see table
\ref{table_variables}. We expect the estimates to be good in $6\leq
D\leq 9$, where we could approach the merger closely, and they are
only bounds (upper or lower, depending on the variable) in higher
dimensions.

It follows that when the spacetime dimension grows the mass and the
entropy of the near-merger solution decrease and their temperature
increases. This is in accordance with the expectation \cite{CritDim}
that in higher dimensions the trends in the behavior of the
(normalized) thermodynamic variables are reversed, see diagram
\ref{fig_phasediagram}. This is best illustrated by the temperature,
which is expected to change its trend in $D \gtrsim 12$. One notes
that while in 6D the limiting temperature along the branch is
smaller than $1$ by about $30 \%$, in 11D it is smaller than $1$ by
less than $1.5\%$. (Of coarse, in this dimension we have not
extended the branch into really asymptotic regime. However, the
shape of the curve $\hT(\lam)$ suggests that the asymptotic value is
not much smaller than the estimate.)

It is interesting to note that our numerical method performs well
for weakly deformed strings in $D\leq11$, but in 12D (and higher) it
fails to converge for any, no matter how small, deformations.
Recalling that the trends in the temperature are expected to change
at about $D \simeq 12$ we tend to attribute this behavior to the
fact that due to limited accuracy our numerics is incapable of
deciding on the sign of $\hT-1$, even if it can determine correctly
other thermodynamic variables.  This probably leads to a conflict
that eventually results in the divergence.

Examining our most deformed strings we provide strong quantitative
evidence that the local geometry in vicinity of their waist is
cone-like. The deviation from the cone geometry is less than ten
percent, which leaves little room for any solution other than the
cone. The cone's presence is further (indirectly) supported by
appearance of the power-law scaling (\ref{scaling}).

The power-law behavior (more precisely -- its onset) of the
characteristic length scale $R_{min}$ is observed and depicted in
figures \ref{fig_Rmin_scaling} and \ref{fig_Rmin_scalingB0}. The
estimates of the corresponding critical exponents are collected in
table \ref{table_gammas}. We used two parameterizations that are not
analytically related, and hence the corresponding $\gamma$'s are not
equal. While the parametrization by $B_0^{-1}$ is not defined beyond
the merger point (along the BHs branch), the parametrization by
$\hL_{hor}$ can be made continuous across this point (though it is
not obvious if it is analytic at the pinch-off); in this sense
$\hL_{hor}$ parametrization is superior.

Proving that  $\hL_{hor}$ is analytic at the pinch-off, will qualify
the parametrization as ``good'' (see footnote 8). In this
case\footnote{If $\hL_{hor}$ won't happen to be analytic at the
merger, the exponents are incomparable.} we can compare the
estimated $\gamma$'s with the prediction of \cite{Kol_on_Choptuik},
where it is proposed that $\gamma_{th} =1/4$ in any $D\leq10$.
Clearly, our estimates differ from $1/4$ and there can be several
reasons for this. First, in general, one should not really expect
for the actual exponents to coincide with those obtained in the
linear analysis of \cite{Kol_on_Choptuik}. Second, our estimates may
be too crude and the more accurate $\gamma$'s will indeed be $1/4$.
While this is possible, we believe that at least in six and in seven
dimensions our estimates cannot be significantly different from the
actual exponents, see figure \ref{fig_Rmin_scaling}. Finally, it is
likely that the truth lies somewhere in the middle, and so are the
actual values of the critical exponents.

In conclusion, we note that there are still many aspects of the
phase diagram \ref{fig_phasediagram} awaiting for a confirmation.
This includes a construction of the black-hole branch in higher than
six dimensions and of the black-string branch in $D>11$. Despite the
progress in understanding the pinch-off geometry reported in this
paper, one still might wish to approach the merger closer and to do
so also form the ``black-hole side". Our current relaxation method
seems inadequate for achieving these goals due to its slow
convergence and instability at the strongest non-uniformity. We believe that in order to address the
mentioned issues our scheme will require major revisions (e.g.
finding a stable multi-grid implementation), or one would probably
have to come up with a completely different approach.

\vspace{0.5cm} \noindent {\bf Acknowledgements}

I would like to thank Barak Kol for useful remarks on the
manuscript, and Carsten Gundlach and Eric Hirschmann for specific
discussions. This research  is supported in  part  by the CIAR
Cosmology and Gravity Program, by NSERC and by the British Columbia
Knowledge Development Fund. The numerical computations were
performed on  UBC's vnfe4 cluster and on Westgrid's Glacier
cluster.

\appendix

\section{Uniform black string in conformal coordinates}
\label{appendix_schw_conf}

The metric of the uniform black string in \Schw coordinates reads
\bea \label{ds_schw} ds^2&=& ds^2_{Schw_d} +dz^2 =
-f(\rho)dt^2+f(\rho)^{-1} d\rho^2+dz^2+\rho^2 d\Omega_{d-2}^2  \non
f(\rho)&=& 1 -\(\rho_0/ \rho\)^{d-3} ,
\eea
where $\rho_0$ designates the horizon location.

 In order  to bring
the $(r,z)$ part of the metric into the conformal form we change the
coordinates such that
\bea
\label{r_rho}
ds^2&=& -f(\rho(r))dt^2+dr^2+dz^2+\rho(r)^2 d\Omega_{d-2}^2  \non
dr/d\rho&=&f^{-1/2}(\rho) \non
r(\rho)&=&r_0 + \int^\rho_{\rho_0} { d\rho' \over \sqrt{ 1-\(\rho_0/ \rho\)^{d-3} } } ,
\eea
where $r_0$ is the integration constant, that designates the
location of the horizon in the conformal gauge (without loss of
generality we choose $r_0=0$.) The integration gives
\bea
 r= \left\{ \begin{array}{llr}
    & \rho_0\sqrt{{\rho\over \rho_0}\({\rho\over \rho_0}-1\)}+ \rho_0\log\(\sqrt{{\rho\over \rho_0}}+ \sqrt{{\rho\over \rho_0}-1} \),
    &\mbox{for}\   d=4;  \label{rho_to_rd4} \\ \\

    &\sqrt{\rho^2-\rho_0^2},   &\mbox{for} \  d = 5 \label{rho_to_rd5}; \\ \\

    &- \rho_0\, { \pi^{1/2} \Gamma({d-4\over d-3}) \over  \Gamma({d-5\over 2(d-3)} ) } + \rho\, _2 F_1 \(-{1\over d-3} ,\half,  {d-4\over d-3}, \[{\rho_0 \over \rho}\]^{d-3} \), &\mbox{for}\   d>5, \label{rho_to_rdg5}
     \end{array}\right.
\eea
where $_2 F_1$ is the hypergeometric function.

Our numerical implementation in the conformal coordinates
requires knowing $\rho(r)$. Explicit inversion of (\ref{rho_to_rd5}) exists only in $d=5$,
in any other dimension the inversion must be achieved numerically.

\section{Details and tests of the numerics}
\label{appendix_numeric}
In this appendix we provide further details of the numerical procedure, verify the constraints and describe how the asymptotic charges are obtained.

Table \ref{table_resolutionsD} summarizes the measures of the domain of integration and the sizes of the coarsest meshes in various dimensions.
In order to minimize the mesh size (and, as a result, the relaxation
time) without compromising accuracy we choose the location of the
asymptotic boundary $r_a$ as close to the origin as possible. To
achieve this, we try several $r_a$'s and select the minimal one for
which the thermodynamic variables $S$ and $T$ undergo less than
$0.5\%$ variation if $r_a$ is increased.
\begin{table}[h!]
\centering \noindent
\begin{tabular}{|c|c|c|c|c|c|}\hline
$D$ &$L/2$ & $r_a$  & low& medium &high \\ \hline\hline
 6  &2.4758 & 7 & $201\times65$ & $401\times129$& $801\times257$ \\  \hline
 7  &1.9884 & 6 & $121\times41$ & $241\times81$ & $481\times161$\\ \hline
 8  &1.6982 & 5 & $101\times35$ & $201\times69$ & $401\times137$\\ \hline
 9  &1.5032 &4.5& $91 \times33$ & $181\times65$ & $361\times129$\\ \hline
 10 &1.3659 & 4 & $91\times33$ & $181\times65$  & $361\times129$  \\ \hline
 11 &1.2566 &3.75& $91 \times33$ & $181\times65$ &  -- \\ \hline
  \end{tabular}
\caption[]{Location of the outer boundary $r_a$  and  sizes of the coarsest meshes
used in various dimensions. $L/2=\pi/k_c$, where $k_c$ is the critical GL wavenumber, see table 1 in \cite{KolSorkin}.}
\label{table_resolutionsD}
\end{table}

\subsection*{Relaxation parameters}

\begin{figure}
\centering
\noindent
\includegraphics[width=8cm]{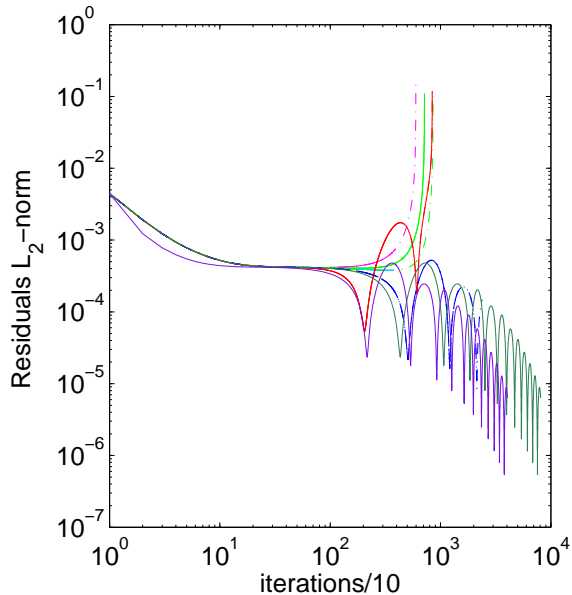}
\caption[]{A typical behavior of the residuals for several sets of
the  parameters $\omega_B$ and $\mu_C$. In some cases the
residuals blow up after finite amount of iterations or show very
slow convergence. In other cases the convergence rate is faster. }
\label{fig_resids}
\end{figure}
The parameters $\omega_B$ and $\mu_C$ are selected after several
trial runs with similar geometry and resolution but with different
parameters. Figure \ref{fig_resids} shows the behavior of the
$L_2$-norm of the residuals in several cases. (The $L_2$-norm of
matrix $M$ on  mesh $[1\dots N_r]\times[1\dots N_z]$ is defined by $
\|M\|_{L_2} \equiv \sqrt{\sum_{i,j} |M_{i,j}|/N_r N_z}$.) While for
some sets of the parameters, the residuals  blow up or converge very
slowly, for other sets the residuals decrease faster. We have not
attempted to locate the ``optimal" parameters that might accelerate
the convergence rate.

The necessity to use under-relaxation for $B$, strong asymptotic
inertia for $C$ and to fine-tune the corresponding parameters into a
narrow range is bewildering.  In addition, the
convergence/divergence of the code is strongly affected by the
initial guess. As described in section \ref{sec_numerics}, for the
relaxation to converge one has to choose the initial guess ``close
enough" to the final solution, otherwise the code diverges.

We tend to attribute this unhappy situation to the fact that $ (1-e^{2\,B-2\,C})/\rho^2$ terms in
 (\ref{Eqs}) have ``wrong sign'' (see also the discussion in \cite{cagedBHsII}).
In order to explain what is wrong with the sign, we linearize around
some solution. Doing so for $\Psi=B$ or $C$ we end up with the
equation of the form
 $\nabla^2 \Psi +m^2 \Psi +\ldots=0$, where $m$ depends on the background solution and the position,  and the ellipsis designate  terms
containing   first derivatives.  Ignoring for a moment these terms we locally get a Helmholtz-type equation. A boundary-value problem  for this type of equation is ill-posed if $m$ is smaller than the mesh box, namely if the tachyon defined by the equation fits into the box and might get exited. This clearly destroys any hopes to relax such an equation numerically.

As for the original set of equations (\ref{Eqs}),  the  finite range
of the fine-tuned parameters $\omega_B$ and $\mu_C$ and the
sensitivity to the initial guess apparently indicates that while
outside of this range the``wrong-sign'' causes divergence, inside
this range the tachyon is suppressed, and the relaxation converges.
We note also that for increasingly non-uniform strings and in higher
dimensions the convergence window shrinks (in 10D, for example,
$\mu_C<0.01$) The sensitivity to specific values of $\omega_B,
\mu_C$ and to the initial guess may explain the failure of the full
multi-grid technique. The method operates on grids of variable size
and this probably causes the excitation of the tachyon, since it may
happen to fit into one of the grid boxes.

Note that at least one of the wrong-sign equations is a direct consequence
of working in a higher dimensional spacetime, $D>4$, that has $O(N+1)$ isometry with $D-3 \geq N \geq 2$. Indeed, in this case the most general metric is given by
\be
\label{SO_N_metric}
ds^2=ds^2_{D-N}(X) + e^{2\,C(X)}\,d\Omega^2_{N}.
\ee
Dimensionally reducing the Einstein-Hilbert action over the
$\mathbb{S}^{N}$ sphere  and varying it to get the equations (see
for example appendix A of \cite{cagedBHsII} where this is elaborated
for $N=D-3$), one finds out that $C$ is governed by 
\be \label{eqnC}
\triangle C - G(g^{\alpha\beta}) e^{-2\,C} +\cdots = 0 , \ee 
where
$g_{\alpha\beta}$ is a metric on $X$, $\triangle \equiv g_X^{-1/2}
\pa_\alpha (g_X^{1/2} g^{\alpha\beta} \pa_\beta)$  and  ellipsis
designate terms  with first order derivatives and second order
derivatives of $g_{\alpha\beta} $. The function $G(g^{\alpha\beta})$ depends on the metric $g^{\alpha\beta}$, 
and it is positive in $D\geq5$ and zero otherwise. The equation
(\ref{eqnC}) has wrong sign in the sense described above, and this
property is invariant under gauge transformations of the manifold
$X$.

\subsection*{Constraints}

A violation of constraints  is inevitable in any numerical solution.
Here, we examine the convergence rate of the constraints $\cU$ and
$\cV$ as a function of the grid-spacings. In addition, we verify
that the (properly normalized) constraints are small.

Usually we solved the equations on the three-grid hierarchy. We
computed $L_2$-norms of both constraints on these grids and checked
that typically the ratio
$(\|\cU\|^{h}-\|\cU\|^{h/2})/(\|\cU\|^{h/2}-\|\cU\|^{h/4}) \gtrsim 3
$ and the similar ratio for $\cV$ is slightly larger than one. This
indicates that while the constraint $\cU$ converges well, the
convergence of $\cV$ is poor. We tend to attribute this behavior of
$\cV$ to a growth of the discretization errors near the horizon due
to the terms containing the singular $1/\sqrt{f}$. We will
demonstrate now that $\cV$ is actually very small everywhere except
in the horizon vicinity.

In order to get an idea of how well the constraints are satisfied,
we define the ``relative constraints'': $ rel ~ \cU \equiv \cU/\sum
|{\rm terms~ in} \, \cU|$ and $rel ~ \cV \equiv \cV/\sum |{\rm
terms~ in}\, \cV|$, where the sum is over the absolute values of all
additive terms forming $\cU$ or $\cV$.
\begin{figure}
\centering
\noindent
\includegraphics[width=14cm,type=eps,ext=.eps,read=.eps]{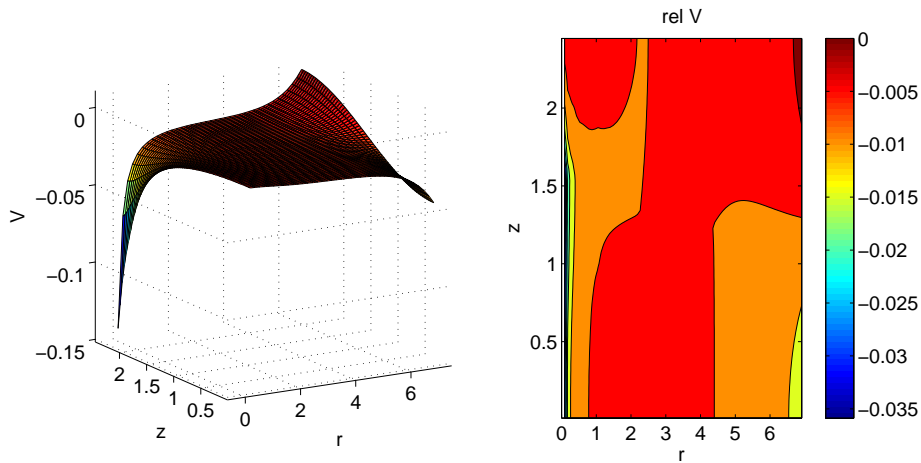}
\caption[]{$\cV $ and $rel ~ \cV$ for a strongly  nonuniform string
($\lam \simeq 3.5$) in 6D. The maximal deviation from zero occurs at
the horizon, near $z=L/2$. Outside this location the constraint is
small. The relative constraint is small too. Altogether we
conclude that the constraint is well satisfied. } \label{fig_conV}
\end{figure}
\begin{figure}
\centering
\noindent
\includegraphics[width=14cm,type=eps,ext=.eps,read=.eps]{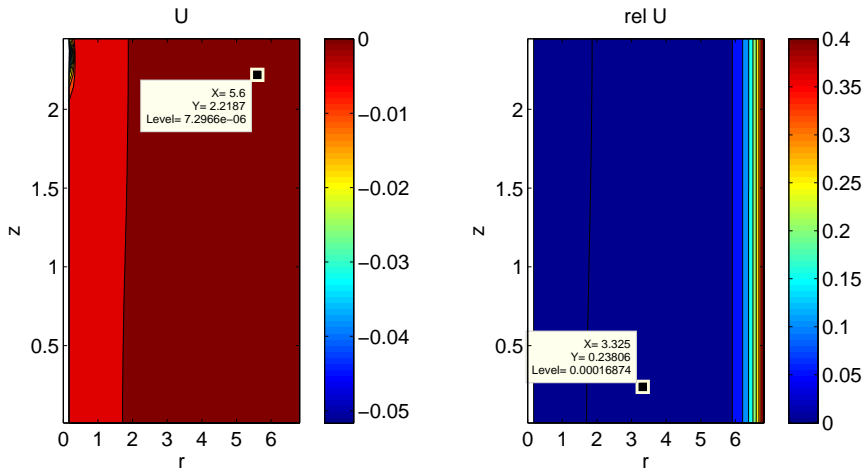}
\caption[]{The $G_r^z \propto \cU$ and $rel ~\cU$ constraints for a
strongly non-uniform string ($\lam \simeq 3.5$) in 6D. The
constraint itself is essentially  small, with the maximum reached
near the horizon. The relative constraint is also very small. The
observed growing towards the outer boundary is caused by smallness
of  $\cU$, see the insertions.  Both plots are complementary and
together they indicate that the constraint is well satisfied. }
\label{fig_conU}
\end{figure}
In figure \ref{fig_conV} we depict  $\cV$ and its relative value.
$\cV$ is maximal near the horizon around $z=L/2$ and it is small
everywhere else. On the other hand the relative constraint is {\it
small} at this corner, but it is large along the remaining section
of the horizon (and its neighborhood), where the $\cV$ itself is
small (typically of order of $10^{-3}-10^{-2}$). Therefore, both
plots are complimentary and together they indicate that the
constraint is well satisfied. The peak that $\cV$ has at the
horizon, see figure, gets sharper at higher resolution, but at the
same time it also moves towards the corner. This is why the $L_2$-norm of
$\cV$ showed poor convergence.

The second constraint $\cU$ and its normalized counterpart are
illustrated in figure \ref{fig_conU}. $\cU$ is  small everywhere and
it consistently diminishes  with increasing grid density.  The relative $\cU$ is very
small as well, except near the outer boundary, where it gets large
simply because $\cU$ is extremely small there and thus the ratio
that forms $rel~ \cU$ becomes of order unity.  In summary, the plots
are complimentary and indicate that the constraint is well
satisfied.

\subsection*{Extraction of the charges}

The mass and the tension can be obtained from the asymptotic
constants (\ref{asymptotics})  using (\ref{asymp_to_charges}). The
constants are extracted by fitting the fields near the outer
boundary with analytic expressions of the form (\ref{asymptotics}).
It turns out that doing the fitting near $r_a$ is unreliable. Hence,
in practice we compute the (averaged along $r_a$) fields and their
r-derivatives, and use this to integrate the fields from $r_a$ to a
larger (typically $\sim 10\, r_a$) distance. The constants are then
extracted from fitting near this boundary. We integrate (\ref{Eqs})
to get $A$ and $C$. In order to find $B$ we solve the first order
equation obtained after the second derivative $\pa_r^2 B$ in
(\ref{Eqs}) is eliminated with the aid of the constraint $G_{zz}$.
(Using the second order equation (\ref{Eqs}) causes blow-up, see
also \cite{Wiseman1}). Of course in the asymptotic equations we drop
the (negligible at large $r$'s) terms containing z-derivatives.

We find that the constant $a$ is essentially robust as it has very
little dependence on the details of the integrating-fitting
procedure. On the other hand, $b$ is very sensitive to these details
and might vary as strong as by $50\%$ when some of them change (see
\cite{cagedBHsII,KudohWiseman1,Wiseman1} for related discussions).
Therefore the mass, which is essentially determined by $a$, is
obtained accurately,  but the tension, which is mainly contributed
by $b$, is not, see (\ref{asymp_to_charges}). In fact we could only
confirm that the relative tension is consistently within the range
$[0,1/(D-3)]$ in  dimensions $D=6,7,8$.

On the other hand, the local horizon quantities $S$ and $T$ are more
accurate since they are measured directly without any additional
manipulations that can enhance numerical errors. Hence, it makes
sense to try computing the mass by integrating the first law
(\ref{1stlaw}) along the solutions branch,
\be
\label{m_1law}
m_{i}(\lam)=m(0)+\int_0^\lam T(\lam')\, \frac{d S}{d\lam'} \, d\lam'.
\ee
We found that the mass computed in this way is close to that
obtained from the asymptotic constants. The agreement improves when
the numerical resolution increases, see figure \ref{fig_mass6D};
this indicates the convergence of the numerics. The discrepancy
between the mass computed in either way is never above $5\%$.

Encouraged by this and by the apparent accuracy of Smarr's formula
(\ref{smarr}), shown in  figure \ref{fig_smarr}, one might hope to
determine the tension without need to invoke the asymptotic
integration-fitting procedure. Unfortunately  this is not the case.
From  (\ref{smarr}) it follows that the tension is given by the
difference of two large numbers,
\be
\label{tauSmarr}
\tau= \((D-3) \,M -(D-2)\, S\, T\)/L
\ee
each of which becomes larger than  $\tau\,L$ by more than an order
of magnitude for $\lam \gtrsim 0.1$. One has to know $M, S $
and $T$ accurately enough to be able to reliably extract $\htau$
from (\ref{tauSmarr}). It follows that in our case $\htau$ computed
from (\ref{tauSmarr}) is very noisy, sometimes with more than $50 \%
$ scattering. In addition, the accuracy worsens in higher
dimensions, partly due to the $D$-dependent factors in
(\ref{tauSmarr}). Consequently, this method does not produce more
accurate estimate of $\tau$.


\end{document}